\documentclass{aa}
\usepackage{graphicx}
\usepackage{txfonts}
\usepackage{natbib}
\usepackage{longtable}

\bibpunct{(}{)}{;}{a}{}{,}  

\begin{document}

\title{Palladium and silver abundances in stars with [Fe/H] $>-2.6$}

\author{
    Xiaoshu Wu \inst{1,2,3}  \and
    Liang Wang \inst{1}       \and
    Jianrong Shi \inst{1}     \and
    Gang Zhao \inst{1}        \and
    Frank Grupp \inst{3}
}

\offprints{G. Zhao; \email{gzhao@nao.cas.cn}}

\institute{
    Key Laboratory of Optical Astronomy,
    National Astronomical Observatories, Chinese Academy of Sciences,
    A20, Datun Road, Chaoyang District, Beijing 100012, China\\
    \email{xswu@nao.cas.cn;gzhao@nao.cas.cn}
    \and
    University of the Chinese Academy of Sciences,
    19A, Yuquan Road, Shijingshan District, Beijing 100049, China
    \and
    Max-Planck-Institut f\"ur Extraterrestrische Physik,
    Giessenbachstrasse, D-85748 Garching, Germany
}

\date{Received; accepted}

\abstract
    {
    Palladium (Pd) and silver (Ag) are the key elements for probing the weak
    component in the rapid neutron-capture process ($r$-process) of stellar
    nucleosynthesis.
    We performed a detailed analysis of the high-resolution and high
    signal-to-noise ratio near-UV spectra from the archive of HIRES on the Keck
    telescope, UVES on the VLT, and HDS on the Subaru Telescope, to determine
    the Pd and Ag abundances of 95 stars.
    This sample covers a wide metallicity range with $-2.6\lesssim$ [Fe/H]
    $\lesssim+0.1$, and most of them are dwarfs.
    The plane-parallel LTE MAFAGS-OS model atmosphere was adopted, and the
    spectral synthesis method was used to derive the Pd and Ag abundances from
    \ion{Pd}{i} $\lambda$ 3404 \AA\ and \ion{Ag}{i} $\lambda$ 3280/3382 \AA\
    lines.
    We found that both elements are enhanced in metal-poor stars, and their
    ratios to iron show flat trends at $-0.6<$ [Fe/H] $<+0.1$.
    The abundance ratios of [Ag/H] and [Pd/H] are well correlated over the whole
    abundance range.
    This implies that Pd and Ag have similar formation mechanisms during the
    Galactic evolution.
}


\maketitle

\section{Introduction}

More than half a century ago, forerunners established the theoretical framework
    for exploring the nucleosynthesis mechanisms.
The nuclides with $Z>30$ are produced by neutron-capture process
    \citep{Cameron1957, Burbidge1957}, including at least two different dominant
    processes -- the rapid one and the slow one.
The slow neutron-capture nucleosynthesis process ($s$-process) accounts for the
    production of around half of the nuclear species beyond the iron-peak
    elements and takes place in a relatively low neutron density environment
    \citep{Kappeler1989, Zhao1990, Busso1999, Kappeler2011}, whereas the rapid
    neutron-capture process ($r$-process) is responsible for the other half and
    occurs in a high neutron density environment \citep{Kratz2007, Sneden2008,
    Farouqi2009, Farouqi2010}.
Besides, the lighter element primary process \citep[LEPP;][]{Travaglio2004,
    Montes2007, Arcones2011} and the $p$-process \citep{Arnould2003} are also
    thought to contribute to the overall stellar abundance patterns.
As a result, the $r$-process always arises in explosive environments that can
    offer enough dense neutrons, such as neutron star mergers
    \citep{Freiburghaus1999, Rosswog1999, Goriely2005, Goriely2011,
    Korobkin2012, Perego2014, Just2015}, neutrino-driven wind
    \citep{Woosley1997, Arcones2011, Wanajo2012}, jets in core-collapse
    supernova \citep{Cameron2001}, massive core-collapse supernova
    \citep{Wasserburg2000, Argast2004}, gamma-ray bursts \citep{McLaughlin2005},
    and low-mass supernova explosion from the collapse in O-Ne-Mg cores
    \citep{Wanajo2003}.

In contrast to this, the $s$-process happens in relatively peaceful environments.
Much research has shown that the $s$-process can be classified into two
    subprocesses.
The main $s$-process, which happens in asymptotic giant branch (AGB) stars with
    masses between 1.3 and 8 $M_\odot$, creates the heavier elements, such as Ba
    \citep[e.g.,][]{Straniero1997, Gallino1998, Busso1999, Karakas2009,
    Cristallo2009, Bisterzo2010}, while the weak $s$-process is associated with
    massive stars with initial mass $>8$ $M_\odot$, and it creates the lighter
    elements, such as Sr \citep{Pignatari2010, Pignatari2013, Frischknecht2012}.
In addition, a third sub-$s$-process, the ``strong'' component, was proposed to
    explain about 50\% of $^{208}$Pb in the Sun \citep{Clayton1967} but was
    later re-interpreted as the outcome of the main $s$-process at low
    metallicity \citep{Gallino1998,Travaglio2001}.

Although several scenarios may be responsible for the $r$-process, their sites
    remain unclear.
Observations suggest that the $r$-process can be divided into two distinct
    components, namely the main $r$-process and the weak $r$-process
    \cite[e.g.,][]{Burris2000, Francois2007, Hansen2011}.
Among the elements with 38 $<$ Z $<$ 50, Pd and Ag are considered as the keys to
    an investigation of the weak $r$-process.
\cite{Crawford1998} reported the first detection of Ag abundances in four
    metal-poor stars and found that $\left<{\rm [Ag/Fe]}\right>\simeq0.22$ dex
    without any trend with [Fe/H].
\cite{Johnson2002} determined Pd and Ag abundances for three metal-poor stars
    and upper limits of Pd for another nine stars, but their correlation is not
    well established owing to the very small number of samples.
Other studies gave Pd or Ag abundances for a few isolated metal-poor stars,
    e.g.,
    CS\,31082-001 \citep{Hill2002},
    CS\,22892-052 \citep{Sneden2003}, and
    HD\,122563 \citep{Honda2006}.

It was not until \cite{Hansen2011} and \cite{Hansen2012} that Pd and Ag
    abundances were analyzed for a large sample of stars.
These authors derived Pd and Ag abundances for 34 dwarfs and 23 giants.
The [Pd/Fe] and [Ag/Fe] show flat trends with metallicity and were compared
    with several tracer elements such as Sr, Y, Zr, Ba, and Eu.
The comparisons have ruled out the weak/main $s$-process or main $r$-process as
    the main formation channels of Pd and Ag.

To study the formation processes of Pd and Ag further in the Galactic chemical
    evolution history, we analyzed the archive near-UV spectra for a large
    sample of stars, from three 8-10m class telescopes.
These samples cover a wide metallicity range ($-3.1\lesssim$ [Fe/H] $\lesssim
    +0.1$) and represent different Galactic populations (thin disk, thick
    disk, halo).
This paper is organized as follows.
Section 2 describes the observations and data reduction; section 3 explains the
    atmosphere models, stellar parameters, and atomic data; section 4 gives the
    chemical abundance and error estimations; and the results are discussed in
    section 5.
In the last section, we summarize our work and present conclusions.

\section{Observation and data reduction}
We searched the archive data of the High Resolution Echelle Spectrometer
    \citep[HIRES;][]{Vogt1994} attached to the Keck\,I Telescope to find the
    high-resolution ultraviolet spectral observations covering the wavelengths
    of $3200\,\AA\le\lambda\le3500\,\AA$, where \ion{Pd}{i} and \ion{Ag}{i}
    lines might exist.
We found 134 stars, of which spectra with high signal-to-noise ratios (S/N) are
    available from Keck\,I/HIRES.
Most of these spectra were taken in the same project as the one that aims to
    determine beryllium abundances with the Be resonance doublet at
    $\lambda\,3130.4/3131.1 \AA$ \citep{Boesgaard2011}.
The typical S/N is above 100 for the \ion{Ag}{i} region and above 120 for
    the \ion{Pd}{i} region.
In addition, these spectra have been taken with the same instrumental
    configuration, and the wavelength resolving power ($\lambda/\Delta\lambda$)
    is $\sim48~000$.
The data reduction was performed using the IRAF\footnote{IRAF is distributed by
    National Optical Astronomy Observatories, operated by the Association of
    Universities for Research in Astronomy, Inc., under contract with the
    National Science Foundation, USA.} {\sc ECHELLE} software package,
    following the standard procedure including bias correction, flat fielding,
    background subtraction, and wavelength calibration.
Whenever possible, one-dimensional spectra of several continuous frames for the
    same star were co-added to obtain a higher S/N and to remove the cosmic-ray
    hints.
The continuum were normalized with polynomials, and the radial velocities were
    corrected by finding the maximum of their cross-correlation functions with
    the solar spectra.
Of the 134 stars, about half (60 stars) have detectable \ion{Pd}{i} line at
    $\lambda\,3404.5\,\AA$ or \ion{Ag}{i} resonance doublet at
    $\lambda\,3280.7/3382.9\,\AA$, all of which are main-sequence or
    slightly-evolved stars, with effective temperatures ($T_{\rm eff}$) ranging
    from 5000\,K to 6350\,K and metallicities ranging from [Fe/H] $=-2.6$ to
    0.05.
    
To enlarge our sample, we searched the archive data of the High Dispersion
    Spectrograph further \citep[HDS;][]{Noguchi2002}, which is mounted on the
    8.2m Subaru Telescope.
We selected seven stars with high S/Ns and clear \ion{Pd}{i} or \ion{Ag}{i}
    lines in the same wavelength region.
All of the stars were observed with StdUb setting and the resolving power is
    $\sim$50,000, comparable to that of HIRES sample.
We reduced their spectra in the same manner as we used with Keck/HIRES data.
This subsample covers a wide metallicity range ($-2.63\le$ [Fe/H] $\le0.02$)
    and contains two metal-poor giants (HD\,6268 and HD\,110184).

In addition, we selected high S/N spectra of 28 stars taken with the Ultraviolet
    and Visual Echelle Spectrograph \citep[UVES;][]{Dekker2000} at ESO VLT 8.2m
    Kueyen telescope.
25 of them were obtained from the ESO Science Archive, and the remaining three were
    taken from the UVES-POP survey \citep{Bagnulo2003}.
This subsample was analyzed in previous research on Be abundances
    \citep{Tan2009}, ensuring a large overlap with the dwarf sample of
    \cite{Hansen2011} and \cite{Hansen2012}, which enables us to make a direct
    comparison of our results with the previous studies.
These spectra were reduced using the {\sc ESO MIDAS}
    package\footnote{http://www.eso.org/sci/software/esomidas/} of version from
    February 2008 in a similar manner to the HIRES and HDS procedures.
The resolving power is $\sim$ 48~000 around 3300 \AA.

\section{Stellar parameters}
\label{sect-para}

\subsection{Effective temperatures}
\label{sect-teff}

Many of our sample stars have been studied well in previous research (e.g.,
    \citet{Chen2000, Reddy2003, Boesgaard2006, Tan2009}; \citet{Boesgaard2011},
    but each uses a part of our entire sample.
Directly mixing the stellar parameters from the literature will cause
    non-negligible biases because these authors adopted different approaches to
    deriving atmospheric parameters ($T_{\rm eff}$, $\log{g}$, [Fe/H], and
    $\xi$).
For instance, \citet{Chen2000} and \citet{Reddy2003} both determined
    $T_{\rm eff}$ from the narrow-band photometric $b-y$ and $c_1$ given by
    \citet{Hauck1998} with the same color-$T_{\rm eff}$ relation
    \citep[i.e.,][]{Alonso1996} and $\log{g}$ from the triangular parallaxes.
While the $T_{\rm eff}$ and $\log{g}$ given by \citet{Boesgaard2006} were
    derived using excitation equilibrium of \ion{Fe}{i} lines and an ionization
    balance between \ion{Fe}{i} and \ion{Fe}{ii}.

Some research has shown systematic deviations between these two methods
    \citep[e.g.][]{Nissen2013}, and the reason can be attributed to the non-LTE
    effect of \ion{Fe}{i} lines \citep[e.g.,][]{Mashonkina2011, Korn2003}.
Recent study by \cite{Lind2012} has shown that the depatures from LTE of
    \ion{Fe}{i} lines result in underestimations by up to 0.4\,dex of
    spectroscopic $\log{g}$ for metal-poor stars \citep[see also][]{Bensby2014}.
Considering many of our samples are nearby stars within 100\,pc and therefore
    have precise {\sc Hipparcos} parallaxes with relative uncertainties less
    than 10\%, we determined the stellar parameters with the former approach.

For 28 stars that we share with \citet{Chen2000,Chen2001} or \citet{Reddy2003},
    we adopted the parameters determined by these studies, in which the narrow
    band photometric colors $b-y$ and $c_1$ from the uvby$\beta$
    catalog\footnote{http://cdsarc.u-strasbg.fr/viz-bin/Cat?II/215}
    \citep{Hauck1998} were used to derive $T_{\rm eff}$ with the calibration of
    \citet{Alonso1996}, and $\log{g}$ were based on the {\sc Hipparcos}
    parallaxes.
A comparison of 23 common stars between \citet{Chen2000} and \citet{Reddy2003}
    shows excellent agreement in the stellar parameters between these two
    datasets, with the mean differences of only $8\pm41$ K, $0.04\pm0.13$ dex,
    and $-0.02\pm0.06$ dex on $T_{\rm eff}$, $\log{g}$, and [Fe/H],
    respectively.

\begin{figure}[htbp]\centering
\includegraphics[width=9cm]{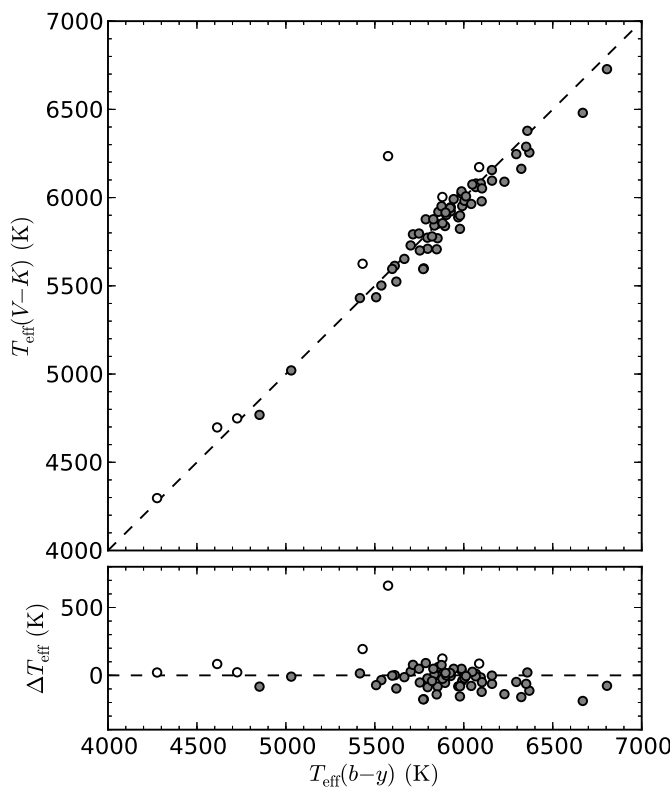}
\caption{
    Comparison of $T_{\rm eff}$ derived from the $b-y$ and $V-K$ colors for 66
    stars.
    For stars with $d>100$\,pc ({\it open circles}), the reddening
    $E(V-K)=2.72E(B-V)$ was taken into account (see text).
}
\label{fig-teff}
\end{figure}

To minimize the systematic bias, for stars not included in \citet{Chen2000} or
    \citet{Reddy2003}, we preferably adopted the photometric colors $b-y$ and
    $c_1$ from \citet{Hauck1998} to derive their $T_{\rm eff}$ with the same
    calibration relation of \citet{Alonso1996}.
This made up another subset of 36 stars.
For the remaining stars in our sample, broad-band colors $V-K$ were used to
    obtain $T_{\rm eff}$ with the calibration relation of \citet{Alonso1996}.
The $V$ magnitudes were taken from the {\sc Hipparcos Catalogue}, and
    $K_{\rm s}$ from the Two Micron All-Sky Survey (2MASS) were converted to $K$
    magnitudes in TCS system according to \citet{Ramirez2004}.
Reddening due to interstellar extinction is only considered for stars with
    distances over 100 pc, and the values were interpolated from the dust map of
    \citet{Schlegel1998} based on COBE and IRAS satellites.
Previous investigations \citep[e.g.,][]{Arce1999} have shown that the $E(B-V)$
    given by this map is slightly overestimated.
We therefore reduced the values according to the formula given by
    \citet{Bonifacio2000} if $E(B-V)>0.10$ and then further by a factor of
    $1-\exp(-|D\sin{b}|/125)$, where $D$ is the distance and $b$ is the Galactic
    latitude of a given star.

Figure \ref{fig-teff} compares the $T_{\rm eff}$ derived from $b-y$ and $V-K$
    for 66 sample stars with the empirical calibration relations of
    \citet{Alonso1996}.
We found satisfactory agreement in the temperatures from these two colors.
The mean difference $\left<T_{\rm eff}(V-K)-T_{\rm eff}(b-y)\right>$ is $-12\pm
    112$\,K, and it drops to $-22\pm75$\,K after excluding the outlier HD\,24289,
    which is probably affected by the large uncertainty ($\sim$37\%) on its
    parallax distance and a relative high reddening value ($E(B-V)\simeq0.12$)
    toward this direction.
The typical errors of $b-y$ (0.003) and $c_1$ (0.007) in \cite{Hauck1998} lead
    to the uncertainties of $\sim25$\,K for our sample stars, and the errors of
    0.10\,dex on [Fe/H] translate into only 5\,K on $\Delta T_{\rm eff}$.
While for stars with $T_{\rm eff}$ derived with $V-K$, the typical uncertainties
    on the color indices and the transformations of $K$ from 2MASS to TCS
    systems are around 0.036 and 0.035\,mag, respectively.
Therefore we estimated the $T_{\rm eff}$ uncertainties to be $\sim$ 80\,K,
    combining with the contributions of $\Delta$[Fe/H]$\simeq0.10$\,dex.
Although the difficulty estimating accurate interstellar reddenings puts
    more uncertainties on $T_{\rm eff}$, this is not the dominant factor for
    the majority of our samples.

\subsection{Surface gravities}

The surface gravities were determined from the fundamental equation

\begin{equation}\label{logg}
\log{g} = \log{g}_\odot + \log\left(\frac{M}{M_\odot}\right)
        + 4\log\left(\frac{T_{\rm eff}}{T_{\rm eff, \odot}}\right)
       + 0.4(M_{\rm bol}-M_{\rm bol, \odot})
\end{equation}

\noindent where $M$ denotes the stellar mass, and $M_{\rm bol}$ is the absolute
    bolometric magnitudes.
We determined the masses by interpolating the Y$^2$ evolutionary tracks
    \citep{Yi2003} with given metal content $Z$ ($\simeq 0.020 \times
    10^{\mathrm{[Fe/H]}}$) and by finding the track that passed through the
    corresponding points of the sample star on the ($T_{\rm eff}$, $L$) plane.
The {\sc Hipparcos} parallaxes \citep{vanLeeuwen2007} were essential for
    determining the absolute magnitude $M_{\rm V}$ and bolometric correction
    with the relation of \cite{Alonso1995}.
According to Formula \ref{logg}, a relative 15\% uncertainty for parallax will
    cause a $\log{g}$ uncertainty of 0.13\,dex and will contribute most to the
    total error of the surface gravity.
By considering the uncertainties on stellar masses and BC \citep[see
    discussion in][]{Tan2009}, we estimated our $\log{g}$ uncertainties to be
    $\pm$0.15\,dex.
Considering that most of our sample stars have parallax uncertainties less than 10\%,
    this is a conservative estimation.

However, this method is not feasible for ten of our sample stars without
    parallaxes.
Alternatively, we adopted the surface gravities by forcing the \ion{Fe}{i} and
    \ion{Fe}{ii} lines that give the same mean iron abundances.
This was done for eight stars (G\,10-4, G\,21-22, G\,24-25, G\,88-10, G\,113-22,
    G\,126-36, G\,130-65, and G\,191-55), and their uncertainties were estimated
    by adjusting $\log{g}$ by an amount that led to an offset of $\Delta$[Fe/H]
    = 0.1\,dex between the mean \ion{Fe}{i} and \ion{Fe}{ii} abundances.
The average \ion{Fe}{i} abundance do not change significantly with the
    variations in $\log{g}$ within $\pm0.2$\,dex, while the abundances derived
    from \ion{Fe}{ii} lines are more sensitive.
We estimated the uncertainties of $\log{g}$ to be $\simeq0.18$\,dex for stars
    with surface gravities determined with the ionization balance method.
There are fewer in subsample than in the total sample, so the errors of Pd or Ag
    abundances caused by the deviations of $\log{g}$ between the above two
    methods do not significantly bias our results.

\subsection{Iron abundances and microturbulences}

For the stars where we determined the parameters, the iron abundances [Fe/H]
    were calculated by measuring the equivalent widths ($W_\lambda$) of $\sim$20
    selected \ion{Fe}{ii} lines, and the microturbulences ($\xi$) were
    determined by canceling out any trend in their individual abundances with
    the equivalent widths.
The calibration relation of $T_{\rm eff}$ given by \citet{Alonso1996} is the
    function of both [Fe/H] and color, therefore the determination of
    $T_{\rm eff}$, $\log{g}$, [Fe/H], and $\xi$ is an iterative procedure.
We searched the literature and took the [Fe/H] values given by
    \citet{Boesgaard2006}, \citet{Boesgaard2011}, \citet{Simmerer2004},
    \citet{Gehren2006}, \citet{Honda2006} and \citet{Ishigaki2012} as the
    initial values.
We found the parameters converged to $\Delta T_{\rm eff}$ within 3\,K and
    $\Delta$ [Fe/H] within 0.01\,dex after two to three iterations for most of
    our sample stars.
The uncertainties of [Fe/H] are estimated by changing the $\log{g}$ by
    $\pm0.15$\,dex and combining the uncertainties of $\sim$0.08\,dex caused by
    the line-by-line scatter.
In summary, the typical uncertainties for the temperatures, surface gravities,
    iron abundances, and microturbulence were estimated to 80\,K, 0.15\,dex,
    0.10\,dex, and 0.2\,km\,s$^{-1}$, respectively.

\subsection{Model atmospheres}

We adopted the one-dimensional, line-blanketed, and local thermodynamic
    equilibrium MAFAGS-OS atmospheric model \citep{Grupp2004,Grupp2009} for
    sample stars and assumed that the mixing-length parameter
    $\alpha_{\rm CM}=0.82$, where the \cite{Canuto1991,Canuto1992} convection
    theory was used.
The iron opacity was calculated based on the improved solar iron
    abundance of \citet{Lodders2009}, and the opacities for metal-poor stars
    with [Fe/H] $<-0.6$ were calculated using $\alpha$-element abundances
    enhanced by 0.4 dex.

\subsection{Comparison with literatures}
\label{sect-comp-para}

\begin{figure}[htbp]\centering
\includegraphics[width=9cm]{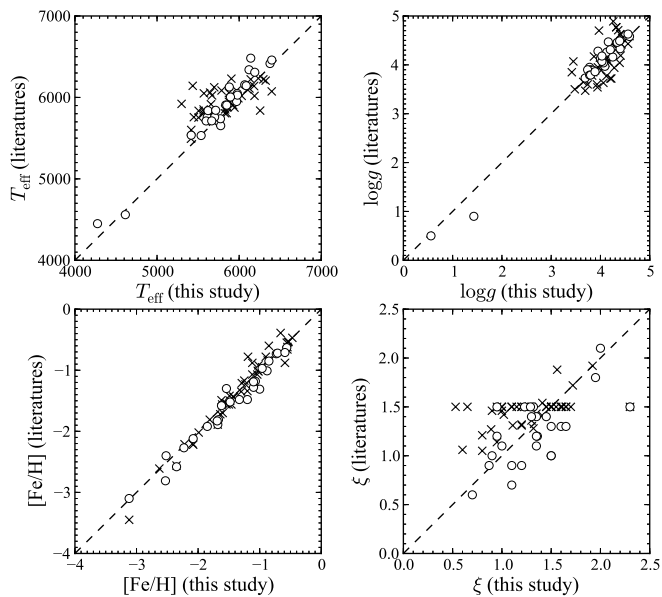}
\caption{
    Comparison of atmospheric parameters ($T_{\rm eff}$, $\log{g}$, [Fe/H] and
    $\xi$) of 25 common stars with \cite{Hansen2012} ({\it open circles}) and 44
    common stars with \cite{Boesgaard2011} ({\it crosses}).
    There are eight overlapped stars in the above two subsamples.
    For the comparison with \cite{Boesgaard2011}, the three stars with
    parameters taken directly from their work were not included.
}
\label{fig-comp-para}
\end{figure}

Figure \ref{fig-comp-para} compared the atmospheric parameters ($T_{\rm eff}$,
    $\log{g}$, [Fe/H] and $\xi$) of common stars with \citet{Hansen2012} and
    \citet{Boesgaard2011}.
\citet{Hansen2012} adopted a similar method to our study to determine the
    stellar parameters.
For example, their $T_{\rm eff}$ were derived with $V-K$ and \cite{Alonso1996}
    relations and $\log{g}$ were based on {\sc Hipparcos} parallaxes.
The mean differences (\citealt{Hansen2012} $-$ this work) are
    $\left<\Delta T_{\rm eff}\right> = 89    \pm 99$\,K,
    $\left<\Delta\log{g}\right>      = 0.04  \pm 0.16$,
    $\left<\Delta{\rm [Fe/H]}\right> = -0.09 \pm 0.12$, and
    $\left<\Delta\xi\right>          = -0.07 \pm 0.26$\,km\,s$^{-1}$
    for 25 common stars.
Despite most of our parameters show good agreements with those of
    \citet{Hansen2012}, temperatures, and/or iron abundances for a few stars
    differ up to $>$200\,K and $\sim$0.3\,dex with theirs, which will cause
    non-negligible influences on Pd and Ag abundances.
These include G\,20-24 ($\Delta T_{\rm eff}=341$\,K), HD\,103723 ($\Delta
    T_{\rm eff}=248$\,K), CD\,$-30$\,18140 ($\Delta T_{\rm eff}=220$\,K),
    HD\,132475 ($\Delta T_{\rm eff}=217$\,K), HD\,111980 ($\Delta{\rm [Fe/H]}=
    -0.31$\,dex), and HD\,106038 ($\Delta{\rm [Fe/H]}=-0.28$\,dex).

We also compared the parameters of 44 shared stars with \citet{Boesgaard2011} in
    Fig. \ref{fig-comp-para}.
These authors derived their stellar parameters with the excitation equilibrium
    and ionization balance method.
The mean differences (\citealt{Boesgaard2011} $-$ this work) are
    $\left<\Delta T_{\rm eff}\right> = 129  \pm 211$\,K,
    $\left<\Delta\log{g}\right>      = 0.04 \pm 0.28$,
    $\left<\Delta{\rm [Fe/H]}\right> = 0.00 \pm 0.13$, and
    $\left<\Delta\xi\right>          = 0.21 \pm 0.26$\,km\,s$^{-1}$,
    which show larger scatters than those of \cite{Hansen2012}.
This is not suprising because the excitation equilibrium and ionization balance
    method relies on accurate \ion{Fe}{i} abundances and is thought to be
    affected by non-LTE effects as mentioned above.

\section{Stellar kinetics}
\label{kinetics}

We derived the Galactic motion velocities ($U_{\rm LSR}$, $V_{\rm LSR}$,
    $W_{\rm LSR}$)\footnote{Here $U$ is defined to be positive towards Galactic
    anticenter.} of the sample stars with the method given by \cite{Johnson1987}.
The radial velocities of HDS and HIRES spectra were measured by finding the
    maximum of the cross-correlation functions with the high-resolution solar
    atlas \citep{Kurucz2005}, and corrected to heliocentric velocities with the
    {\sc rvcorrect} task of the IRAF {\sc astutil} package.
The uncertainties of radial velocities were estimated by quadratically summing
    the drifts of ThAr line positions during an observing night, the R.M.S. of
    ThAr line centroid fitting, and the R.M.S. of measured radial velocities for
    every echelle order.
The typical radial velocity uncertainties for HDS spectra are estimated to be
    around 0.5\,km\,s$^{-1}$ and larger ($\sim$2\,km\,s$^{-1}$) for HIRES
    spectra, which are dominated by larger drifts ($\sim$4\,pixels) of ThAr
    lines in one night.
For stars with UVES spectra, we adopted the values listed in the
    SIMBAD database directly.
The parallax and proper motion data were taken from the new reduction of
    {\sc Hipparcos Catalogue} \citep{vanLeeuwen2007} or the {\sc Tycho-2
    Catalogue} \citep{Hog2000}.
For sample stars without {\sc Hipparcos} parallaxes, we inversed Formula
    \ref{logg} with the spectroscopic $T_{\rm eff}$ and $\log{g}$ that were
    derived in Sect. \ref{sect-para}, together with the basic relation
    $M_{\rm bol} = V_{\rm mag} + BC - 5\log{d} + 5 - A_{\rm V}$ to find their
    distances.

\begin{figure}[htbp]\centering
\includegraphics[width=9cm]{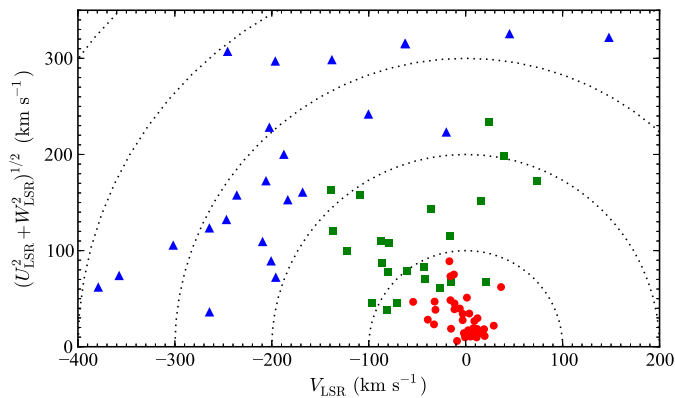}
\caption{
    Toomre diagram of sample stars in this work.
    Stars with the highest probability of belonging to the thin disk, the thick
    disk, and the halo are plotted with {\it red dots}, {\it green squares},
    and {\it blue triangles}, respectively.
    Dotted lines represent total velocities of $|v|_{\rm LSR}=\sqrt{U^2_{\rm
    LSR} + V^2_{\rm LSR} + W^2_{\rm LSR}}$ in steps of 100 km\,s$^{-1}$.
}
\label{fig-toomre}
\end{figure}

We adopted the solar motion of $(U,V,W)_\odot=(-10.00\pm0.36, +5.25\pm0.62,
    +7.17\pm0.38$)\,km\,s$^{-1}$ \citep{Dehnen1998} to correct the $(U, V, W)$
    components of the sample stars to the local standard of rest (LSR).
This allowed us to calculate the relative probabilities for the
    thick-disk-to-thin-disk (TD/D) and thick-disk-to-halo (TD/H) memberships of
    each star with the pure kinematic method proposed by \cite{Bensby2003}.
We emphasize that it is hard to clarify whether a star in the solar neighborhood
    belongs to the thin or the thick disk, and some other studies used chemical
    criteria such as [Fe/H] and [$\alpha$/Fe] \citep[e.g.,][]{Fuhrmann1998,
    Navarro2011}.
However, we simply divided our sample stars into three groups, thin disk, thick
    disk, and halo stars, according to whichever population had the highest
    probability based on the kinematic properties.
We manually assigned four stars (HD\,122563, HD\,84937, HD\,6268, and G\,88-10)
    with [Fe/H] $<-1.5$ that are classified as thin disk stars by the kinetic
    method as halo stars.
Previous studies have shown that all of them are enhanced in magnesium with
    [Mg/Fe] $> 0.49$ \citep[e.g.,][]{Gratton2003,Honda2004,Boesgaard2011}.
Figure \ref{fig-toomre} shows the Toomre diagram of our sample stars, and the
    radial velocities, prallaxes, proper motions, and Galactic velocities are
    listed in the online Table \ref{tab-kinetics}.

\onllongtab{
    \footnotesize{
    \begin{longtable}{lrrrrrrrrrc}
        \caption{Kinetic properties and stellar populations.
            The sample stars are organized into three subcategories, UVES, HDS,
            and HIRES, based the instruments (see online Table
            \ref{tab-abundance}).}
        \label{tab-kinetics}\\
            \hline
            \hline
            Star & $V_{\rm helio}$& $\pi$ & $\mu(\alpha)\cos{\delta}$ & $\mu(\delta)$& $U_{\rm LSR}$ & $V_{\rm LSR}$ & $W_{\rm LSR}$ & Population\\
                 & (km\,s$^{-1}$) &(mas) & (mas/yr)                  & (mas/yr)     & (km\,s$^{-1}$)& (km\,s$^{-1}$)& (km\,s$^{-1}$)&           \\
            \hline
            \endfirsthead
            \caption{Continued.}\\
            \hline
            Star & $V_{\rm helio}$& $\pi$ & $\mu(\alpha)\cos{\delta}$ & $\mu(\delta)$& $U_{\rm LSR}$ & $V_{\rm LSR}$ & $W_{\rm LSR}$ & Population\\
                 & (km\,s$^{-1}$) &(mas) & (mas/yr)                  & (mas/yr)     & (km\,s$^{-1}$)& (km\,s$^{-1}$)& (km\,s$^{-1}$)&           \\
            \hline
            \endhead
            \hline
            \endfoot
            \hline
            \endlastfoot
CD\,$-$30 18140   &    16.8& $  7.16\pm1.45$ & $  -60.68\pm1.80$& $ -319.04\pm1.11$ &  $-$71.50& $-$195.90&  $-$11.20 &halo  \\
CD\,$-$57 1633    &   260.7& $  9.91\pm0.88$ & $  -94.08\pm0.89$& $  689.02\pm1.08$ &    306.30& $-$245.60&  $-$24.90 &halo  \\
G\,13-9           &    57.9& $  4.87\pm1.51$ & $ -278.13\pm1.67$& $ -228.08\pm0.91$ &     93.10& $-$264.40&  $-$81.40 &halo  \\
G\,20-24          &    34.4& $  5.29\pm2.52$ & $ -187.66\pm2.29$& $ -233.10\pm1.99$ & $-$160.30& $-$206.00&     64.30 &halo  \\
G\,183-11         &$-$242.7& $  9.38\pm3.43$ & $ -222.94\pm2.31$& $ -352.20\pm2.45$ &  $-$53.50& $-$379.00&  $-$31.60 &halo  \\
HD\,61421         &  $-$3.2& $284.56\pm1.26$ & $ -714.59\pm2.06$& $-1036.80\pm1.15$ &  $-$12.60&      6.97&      6.43 &thin  \\
HD\,76932         &   119.4& $ 47.54\pm0.31$ & $  244.14\pm0.23$& $  213.94\pm0.15$ &     37.70&  $-$86.20&     78.20 &thick \\
HD\,84937         & $-$15.0& $ 13.74\pm0.78$ & $  373.05\pm0.91$& $ -774.38\pm0.33$ &  $-$18.14&     11.93&   $-$3.52 &halo  \\
HD\,97320         &    53.2& $ 18.36\pm0.58$ & $  160.06\pm0.61$& $ -201.30\pm0.54$ &  $-$83.50&  $-$16.80&  $-$30.50 &thin  \\
HD\,97916         &    61.1& $  8.95\pm1.14$ & $  208.21\pm0.99$& $   -8.80\pm0.96$ & $-$117.70&     15.90&     96.10 &thick \\
HD\,103723        &   168.3& $  5.84\pm1.38$ & $ -234.34\pm1.05$& $  -56.82\pm0.72$ &     65.80& $-$200.70&     60.50 &halo  \\
HD\,106038        &    99.4& $  9.98\pm1.57$ & $ -216.24\pm1.48$& $ -439.18\pm1.02$ &  $-$25.20& $-$264.30&     26.20 &halo  \\
HD\,111980        &   155.0& $ 10.50\pm1.26$ & $  299.64\pm0.96$& $ -794.83\pm0.63$ & $-$277.90& $-$196.40& $-$105.50 &halo  \\
HD\,113679        &   157.8& $  8.75\pm1.38$ & $ -386.94\pm1.23$& $ -145.79\pm0.74$ &    105.80& $-$301.70&      0.10 &halo  \\
HD\,121004        &   245.3& $ 16.70\pm1.24$ & $ -483.51\pm1.47$& $    8.54\pm0.83$ &  $-$74.90& $-$246.80&    109.20 &halo  \\
HD\,122196        & $-$26.4& $ 10.46\pm1.07$ & $ -453.33\pm0.89$& $  -81.62\pm0.83$ &    160.90& $-$139.40&     23.50 &thick \\
HD\,122563        & $-$26.6& $  4.22\pm0.35$ & $ -189.86\pm0.27$& $  -69.67\pm0.19$ &      0.84&      6.91&  $-$17.04 &halo  \\
HD\,126681        & $-$45.6& $ 21.04\pm1.12$ & $  -70.59\pm1.24$& $ -311.72\pm0.72$ &     12.30&  $-$42.20&  $-$68.90 &thick \\
HD\,132475        &   176.5& $ 10.23\pm0.84$ & $ -558.49\pm0.85$& $ -500.37\pm0.68$ &  $-$47.00& $-$357.50&     57.40 &halo  \\
HD\,140283        &$-$169.0& $ 17.16\pm0.68$ & $-1114.93\pm0.62$& $ -304.36\pm0.74$ &  $-$30.00&    147.80& $-$320.50 &halo  \\
HD\,160617        &   100.4& $  9.09\pm1.31$ & $  -61.77\pm1.55$& $ -395.70\pm1.10$ &  $-$68.20& $-$209.50&  $-$85.70 &halo  \\
HD\,166913        & $-$48.6& $ 15.80\pm0.91$ & $ -260.62\pm0.93$& $ -125.75\pm0.68$ &     35.00&  $-$43.30&     75.40 &thick \\
HD\,175179        &    21.7& $ 14.59\pm1.29$ & $ -133.71\pm1.43$& $ -431.46\pm1.11$ & $-$117.70& $-$137.30&  $-$25.60 &thick \\
HD\,188510        &$-$192.5& $ 26.71\pm1.08$ & $  -38.11\pm1.03$& $  287.81\pm1.13$ &    141.10& $-$109.40&     71.60 &thick \\
HD\,189558        & $-$12.9& $ 15.39\pm0.81$ & $ -308.37\pm0.85$& $ -365.10\pm0.86$ &  $-$85.90& $-$122.20&     50.10 &thick \\
HD\,195633        & $-$45.8& $ 10.07\pm0.84$ & $   73.75\pm0.93$& $   22.59\pm0.82$ &     48.30&  $-$15.70&   $-$3.50 &thin  \\
HD\,205650        &$-$105.5& $ 18.14\pm0.96$ & $  342.91\pm0.93$& $ -208.50\pm0.55$ &    106.40&  $-$79.50&     17.30 &thick \\
HD\,298986        &   198.2& $  6.61\pm1.41$ & $  408.70\pm1.28$& $    4.58\pm1.15$ & $-$250.10& $-$138.00&    163.30 &halo  \\
\hline
G\,21-22          &  60.1  &    $\cdots$     & $ -168.90\pm1.50$& $ -446.20\pm1.50$\tablefootmark{a} &  $-$61.27&     36.47&     10.06 &thin  \\
HD\,6268          &   40.5 & $  0.75\pm0.78$ & $  -30.74\pm0.83$& $  -35.10\pm0.80$ &   $-$9.03&      3.70&  $-$33.31 &halo  \\
HD\,94028         &   66.1 & $ 21.11\pm0.92$ & $ -262.06\pm1.02$& $ -456.84\pm0.63$ &     13.82&  $-$15.17&     65.43 &thick \\
HD\,110184        &  140.2 & $  0.76\pm0.84$ & $  -14.14\pm0.74$& $   -9.19\pm0.55$ &  $-$28.54&  $-$36.00&    139.90 &thick \\
HD\,198390        &    5.0 & $ 32.66\pm0.41$ & $   52.94\pm0.38$& $   96.82\pm0.23$ &  $-$12.44&      9.32&      5.52 &thin  \\
HD\,201891        & $-$43.9& $ 29.10\pm0.64$ & $ -122.95\pm0.50$& $ -899.21\pm0.39$ &      6.18&  $-$32.64&     22.44 &thin  \\
HD\,220242        &    7.1 & $ 14.30\pm0.72$ & $  -94.67\pm0.51$& $  -85.17\pm0.44$ &   $-$9.07&     11.23&      3.36 &thin  \\
\hline
BD\,+04 4551      &$-$117.4& $  1.61\pm2.66$ & $   -8.52\pm2.42$& $   77.87\pm1.38$ &     56.30&  $-$79.90&     53.47 &thick \\
BD\,+07 4841      &$-$234.3& $  5.06\pm2.51$ & $  284.87\pm2.73$& $ -102.20\pm1.77$ &     48.69& $-$168.21&    153.37 &halo  \\
BD\,+17 4708      &$-$286.3& $  8.21\pm1.26$ & $  511.75\pm1.34$& $   59.91\pm1.17$ &     42.31& $-$236.09&    152.09 &halo  \\
BD\,+19 1185      &$-$190.8& $ 16.81\pm2.04$ & $  664.05\pm2.02$& $ -622.42\pm1.20$ & $-$197.74&     39.09&     11.84 &thick \\
BD\,+21 607       &   339.9& $ 11.40\pm1.22$ & $  426.35\pm1.23$& $ -301.49\pm0.95$ &    306.31&     45.24& $-$110.73 &halo  \\
BD\,+22 396       & $-$23.5& $ 10.59\pm1.33$ & $   55.79\pm1.43$& $ -359.31\pm1.04$ &  $-$28.02&   $-$3.10&     19.81 &thin  \\
BD\,+37 1458      &   242.8& $  6.46\pm1.31$ & $   71.30\pm1.36$& $ -352.80\pm0.76$ &    228.55&     24.11&     48.54 &thick \\
BD\,+51 1696      &    65.3& $ 12.85\pm1.33$ & $ -870.07\pm1.08$& $ -543.76\pm0.89$ &     14.61&     21.41&     65.55 &thick \\
BD\,$-$01 306     &    28.1& $ 15.87\pm1.23$ & $  994.75\pm1.37$& $  -79.73\pm1.02$ &      4.80&      9.23&  $-$16.39 &thin  \\
BD\,$-$17 484     &   235.3& $  5.75\pm1.56$ & $  400.23\pm1.31$& $  -57.65\pm1.33$ &     87.33&  $-$19.99& $-$205.64 &halo  \\
G\,10-4           &    63.7& $ 15.69\pm2.75$ & $ -567.16\pm3.62$& $ -509.01\pm2.06$ &      1.86&  $-$26.16&     61.33 &thick \\
G\,24-25          &$-$311.9&    $\cdots$     & $  141.60\pm1.40$& $ -144.30\pm1.40$\tablefootmark{a} &    186.04& $-$202.55&    132.43 &halo  \\
G\,63-46          & $-$25.5& $  8.62\pm1.43$ & $  111.35\pm1.39$& $ -286.45\pm1.08$ &   $-$2.40&      7.48&  $-$17.07 &thin  \\
G\,74-5           &    27.4& $ 18.45\pm1.24$ & $  289.75\pm1.29$& $ -265.67\pm1.04$ &      8.98&     19.58&   $-$6.61 &thin  \\
G\,88-10          &    84.1&    $\cdots$     & $  -18.47\pm7.97$& $ -254.93\pm4.40$ &     69.35&  $-$12.15&     28.94 &halo  \\
G\,113-22         &    56.2&    $\cdots$     & $  224.40\pm0.90$& $ -150.40\pm0.90$\tablefootmark{a} &     28.75&  $-$31.19&     25.53 &thin  \\
G\,126-36         & $-$87.1&    $\cdots$     & $ -103.50\pm1.10$& $ -241.00\pm1.10$\tablefootmark{a} &     10.61&  $-$70.81&     44.40 &thick \\
G\,130-65         &$-$270.1&    $\cdots$     & $   57.00\pm1.60$& $ -249.10\pm1.60$\tablefootmark{a} &  $-$97.86& $-$187.56&    174.71 &halo  \\
G\,153-21         & $-$63.4& $ 11.92\pm1.69$ & $ -230.94\pm1.94$& $   42.10\pm1.49$ &     43.25&      1.28&  $-$27.15 &thin  \\
G\,180-24         &$-$151.4& $  8.41\pm1.02$ & $ -194.65\pm1.03$& $ -365.60\pm1.01$ &     29.08&  $-$87.48& $-$106.00 &thick \\
G\,188-22         & $-$93.2& $  9.03\pm1.68$ & $ -237.88\pm1.42$& $ -159.34\pm1.33$ &      5.78&  $-$81.32&     38.00 &thick \\
G\,191-55         &$-$257.2&    $\cdots$     & $  191.10\pm2.90$& $ -110.00\pm2.90$\tablefootmark{a} & $-$233.16& $-$100.33&  $-$65.16 &halo  \\
G\,192-43         &   190.5& $  4.45\pm1.90$ & $   -8.05\pm1.50$& $ -476.62\pm1.37$ &    152.53&     73.06&     79.79 &thick \\
HD\,14877         &    33.4& $ 12.76\pm1.16$ & $  136.34\pm1.25$& $    5.70\pm0.90$ &     13.52&     18.97&  $-$12.26 &thin  \\
HD\,22521         & $-$37.4& $ 24.78\pm1.15$ & $ -196.14\pm1.17$& $ -132.69\pm0.98$ &  $-$42.85&  $-$11.52&     13.88 &thin  \\
HD\,24289         &   129.1& $  4.74\pm1.74$ & $  271.32\pm1.98$& $    6.84\pm1.78$ &     85.10&  $-$16.18&  $-$77.48 &thick \\
HD\,24421         & $-$32.8& $ 26.38\pm0.54$ & $ -119.17\pm0.73$& $  108.36\pm0.58$ &  $-$38.17&  $-$11.58&      7.76 &thin  \\
HD\,25173         &    34.8& $ 18.53\pm0.42$ & $  167.95\pm0.29$& $ -308.55\pm0.42$ &     13.46&     28.94&     17.30 &thin  \\
HD\,26421         & $-$65.4& $ 13.32\pm0.92$ & $  119.37\pm1.00$& $  100.70\pm0.83$ &  $-$70.95&  $-$15.69&     18.44 &thin  \\
HD\,28620         &    21.3& $ 22.96\pm0.59$ & $   30.85\pm0.53$& $   -7.84\pm0.37$ &     10.34&     10.95&      4.39 &thin  \\
HD\,30743         &  $-$2.8& $ 29.58\pm0.49$ & $ -120.83\pm0.47$& $ -172.32\pm0.36$ &  $-$12.07&      6.52&      8.72 &thin  \\
HD\,31128         &   111.9& $ 15.00\pm1.13$ & $  165.55\pm0.81$& $  -27.75\pm1.25$ &     50.22&  $-$60.85&  $-$60.11 &thick \\
HD\,33632         &  $-$0.6& $ 38.29\pm0.55$ & $ -145.00\pm0.50$& $ -135.14\pm0.27$ &  $-$10.67&      5.12&      7.16 &thin  \\
HD\,54717         &     3.6& $ 20.75\pm0.62$ & $ -112.06\pm0.95$& $   -8.52\pm0.55$ &   $-$6.57&      4.99&      8.28 &thin  \\
HD\,63333         & $-$10.2& $ 23.21\pm0.47$ & $ -123.32\pm0.53$& $  -21.93\pm0.45$ &  $-$19.01&      7.87&      3.15 &thin  \\
HD\,68284         &    63.8& $ 13.14\pm0.88$ & $  -10.74\pm1.13$& $   51.46\pm0.69$ &     37.07&  $-$32.05&     28.81 &thin  \\
HD\,80218         & $-$12.0& $ 24.60\pm0.58$ & $ -132.60\pm0.55$& $ -134.26\pm0.26$ &  $-$17.77&     10.12&   $-$0.67 &thin  \\
HD\,89125         &    38.5& $ 43.85\pm0.36$ & $ -414.15\pm0.37$& $  -97.66\pm0.19$ &      9.01&   $-$6.09&     38.71 &thin  \\
HD\,91638         &  $-$4.2& $ 28.69\pm0.53$ & $   -0.82\pm0.50$& $ -161.07\pm0.49$ &  $-$11.01&      8.11&      4.14 &thin  \\
HD\,91889         &  $-$5.4& $ 39.88\pm0.37$ & $  268.46\pm0.30$& $ -672.57\pm0.29$ &  $-$10.89&      9.34&      3.75 &thin  \\
HD\,94835         &     9.6& $ 20.32\pm0.66$ & $ -151.24\pm0.77$& $ -215.57\pm0.49$ &   $-$6.61&      2.53&     15.79 &thin  \\
HD\,100180        &  $-$3.8& $ 42.87\pm1.22$ & $ -329.26\pm1.28$& $ -190.01\pm0.98$ &  $-$10.63&      6.52&      3.63 &thin  \\
HD\,104056        & $-$21.8& $ 13.64\pm1.03$ & $  111.11\pm0.91$& $ -177.14\pm0.63$ &   $-$8.07&     17.36&  $-$10.87 &thin  \\
HD\,109303        &    24.0& $ 11.30\pm0.76$ & $  -97.95\pm0.61$& $  -47.76\pm0.51$ &   $-$3.99&     12.18&     29.41 &thin  \\
HD\,118244        & $-$17.0& $ 21.60\pm0.79$ & $ -248.95\pm0.91$& $  128.71\pm0.51$ &   $-$6.76&      4.66&   $-$9.54 &thin  \\
HD\,134439        &   310.2& $ 34.65\pm1.28$ & $ -997.47\pm1.20$& $-3543.55\pm1.03$ & $-$255.28&  $-$62.10&    184.85 &halo  \\
HD\,134440        &   310.9& $ 35.14\pm1.48$ & $ -999.75\pm1.29$& $-3542.60\pm1.13$ & $-$255.91&  $-$62.56&    184.94 &halo  \\
HD\,186379        &  $-$6.7& $ 22.53\pm0.60$ & $   86.96\pm0.47$& $ -270.98\pm0.52$ &   $-$6.71&   $-$0.65&      7.07 &thin  \\
HD\,194598        &$-$246.4& $ 17.00\pm0.83$ & $  117.25\pm0.83$& $ -551.20\pm0.91$ &    133.00& $-$183.50&     75.55 &halo  \\
HD\,200580        &  $-$6.2& $ 19.27\pm0.99$ & $ -273.99\pm1.03$& $ -371.70\pm0.47$ &   $-$6.72&      0.80&     10.06 &thin  \\
HD\,202884        &  $-$0.2& $ 23.92\pm0.76$ & $  143.76\pm0.90$& $  -41.31\pm0.56$ &   $-$9.86&      5.03&      7.27 &thin  \\
HD\,204712        & $-$25.2& $ 14.56\pm0.66$ & $ -108.73\pm0.75$& $ -142.88\pm0.44$ &   $-$0.62&  $-$15.13&     18.74 &thin  \\
HD\,209320        & $-$49.6& $ 14.18\pm0.87$ & $   19.23\pm0.79$& $   15.35\pm1.07$ &   $-$2.17&  $-$39.08&     28.09 &thin  \\
HD\,209858        &  $-$1.8& $ 18.00\pm0.76$ & $   37.80\pm0.72$& $ -193.41\pm0.80$ &   $-$9.84&      3.57&      7.80 &thin  \\
HD\,215442        &  $-$7.3& $ 13.75\pm0.61$ & $  127.89\pm0.44$& $   75.28\pm0.43$ &  $-$11.67&   $-$1.86&      8.02 &thin  \\
HD\,241253        & $-$15.0& $  8.66\pm1.77$ & $  269.46\pm2.21$& $  -70.08\pm1.23$ &  $-$23.69&      8.96&     12.33 &thin  \\
HD\,247297        &    38.8& $  4.61\pm1.44$ & $   64.07\pm1.65$& $ -189.15\pm1.02$ &     27.56&   $-$3.10&      2.16 &thin  \\
HD\,345957        &$-$115.0& $ 10.42\pm1.14$ & $ -171.60\pm0.56$& $   63.37\pm0.68$ &     41.53&  $-$97.09&     17.70 &thick \\
Ross\,390         &    80.9& $  7.10\pm2.45$ & $  423.11\pm1.74$& $ -482.21\pm1.61$ &     44.55&  $-$54.20&     13.96 &thin  \\
Ross\,797         &    22.6& $  3.02\pm2.38$ & $   76.68\pm1.78$& $ -433.46\pm1.91$ &      6.00&   $-$8.91&   $-$0.40 &thin  \\
                               
    \end{longtable}
    \tablefoot{
    \tablefoottext{a}{Stars with distances derived with spectroscopic
    $\log{g}$, and proper motions were taken from the {\sc Tycho-2 Catalogue}
    (see section \ref{kinetics}).}
    For other stars, parallaxes and proper motions were taken from the new
    reduction of {\sc Hipparcos Catalouge} \citep{vanLeeuwen2007}.
    }
}
}

\onllongtab{
    \footnotesize{
    \begin{longtable}{llcccclrrrr}
        \caption{Stellar parameters and Pd and Ag abundances}
        \label{tab-abundance}\\
            \hline
            \hline
            Star & Instrument & $T_{\rm eff}$ & $\log{g}$ & [Fe/H] & $\xi$ & $T_{\rm eff}$ method & [Pd/Fe] & [Ag/Fe] & [Ag/Fe] & [Ag/Fe]\\
                 &            &  (K)          &           &        & (km\,s$^{-1}$)&&         &  (3280\,\AA)&  (3382\,\AA)&  mean  \\
            \hline
            \endfirsthead
            \caption{Continued.}\\
            \hline
            Star & Instrument & $T_{\rm eff}$ & $\log{g}$ & [Fe/H] & $\xi$ & $T_{\rm eff}$ method & [Pd/Fe] & [Ag/Fe] & [Ag/Fe] & [Ag/Fe]\\
                 &            &  (K)          &           &        & (km\,s$^{-1}$)&&         &  (3280\,\AA)&  (3382\,\AA)&  mean  \\
            \hline
            \endhead
            \hline
            \endfoot
            \hline
            \endlastfoot
CD\,$-$30 18140\tablefootmark{a}   &VLT/UVES                  &6120 &4.00 &$-$1.85 &1.50&$V-K$ &    0.28& $<$ 0.74& $<$ 0.74& $<$ 0.74\\
CD\,$-$57 1633                     &VLT/UVES                  &5836 &4.13 &$-$0.88 &1.35&$V-K$ &    0.01&     0.09&     0.03&     0.06\\
G\,13-9                            &VLT/UVES                  &6376 &3.78 &$-$2.23 &1.35&$V-K$ & $<$0.81& $<$ 1.23& $\cdots$& $<$ 1.23\\
G\,20-24                           &VLT/UVES                  &6141 &4.16 &$-$1.68 &0.95&$V-K$ & $<$0.24& $\cdots$& $\cdots$& $\cdots$\\
G\,183-11                          &VLT/UVES                  &6190 &4.09 &$-$2.08 &1.50&$V-K$ & $<$0.40& $<$ 0.64& $<$ 0.28& $<$ 0.28\\
HD\,61421                          &VLT/UVES\tablefootmark{c} &6668 &3.88 &$-$0.03 &1.95&$b-y$ &    0.06& $<$ 0.41& $<$ 0.19& $<$ 0.19\\
HD\,76932                          &VLT/UVES                  &5849 &4.05 &$-$0.96 &1.60&$b-y$ &    0.13&     0.25&     0.23&     0.24\\
HD\,84937                          &VLT/UVES\tablefootmark{c} &6323 &4.02 &$-$2.09 &1.70&$b-y$ & $<$0.45& $<$ 0.74& $<$ 0.84& $<$ 0.74\\
HD\,97320                          &VLT/UVES                  &5991 &4.14 &$-$1.11 &1.35&$b-y$ &    0.21&     0.33&     0.18&     0.26\\
HD\,97916                          &VLT/UVES                  &6445 &4.16 &$-$0.88 &1.50&$b-y$\tablefootmark{1}&$<$ 0.13& $\cdots$& $\cdots$& $\cdots$\\
HD\,103723                         &VLT/UVES                  &5880 &3.95 &$-$0.85 &1.32&$b-y$ &    0.05&     0.08&     0.08&     0.08\\
HD\,106038                         &VLT/UVES                  &5969 &4.40 &$-$1.20 &1.00&$b-y$ &    0.09&     0.19&     0.15&     0.17\\
HD\,111980                         &VLT/UVES                  &5775 &3.80 &$-$1.00 &0.95&$b-y$ &    0.04&     0.11&     0.17&     0.14\\
HD\,113679                         &VLT/UVES                  &5612 &4.06 &$-$0.56 &1.10&$b-y$ & $-$0.10&     0.06&  $-$0.12&  $-$0.03\\
HD\,121004                         &VLT/UVES                  &5598 &4.34 &$-$0.73 &1.10&$b-y$ &    0.08&     0.11&     0.05&     0.08\\
HD\,122196                         &VLT/UVES                  &5978 &3.85 &$-$1.67 &1.36&$b-y$ &    0.15& $<$ 0.62& $<$ 0.32& $<$ 0.32\\
HD\,122563                         &VLT/UVES\tablefootmark{c} &4614 &1.43 &$-$2.53 &1.95&$b-y$ &$<$ 0.28& $<$ 0.20& $<$ 0.20& $<$ 0.20\\
HD\,126681                         &VLT/UVES                  &5537 &4.59 &$-$1.10 &0.70&$b-y$ &    0.33&     0.26&     0.14&     0.20\\
HD\,132475                         &VLT/UVES                  &5621 &3.74 &$-$1.48 &1.23&$b-y$ &    0.18&     0.07& $<$ 0.25&     0.07\\
HD\,140283                         &VLT/UVES                  &5772 &3.69 &$-$2.35 &1.65&$b-y$ &$<$ 0.45& $<$ 0.64& $<$ 0.84& $<$ 0.64\\
HD\,160617                         &VLT/UVES                  &5979 &3.78 &$-$1.69 &1.50&$b-y$ &    0.34& $<$ 0.37& $\cdots$& $<$ 0.37\\
HD\,166913                         &VLT/UVES                  &6068 &4.03 &$-$1.54 &1.29&$b-y$ &    0.29& $<$ 0.35& $<$ 0.38& $<$ 0.35\\
HD\,175179                         &VLT/UVES                  &5701 &4.25 &$-$0.71 &1.20&$b-y$ &    0.00&     0.15&     0.15&     0.15\\
HD\,188510                         &VLT/UVES                  &5416 &4.56 &$-$1.62 &0.90&$b-y$ &    0.25&     0.24&     0.30&     0.27\\
HD\,189558                         &VLT/UVES                  &5666 &3.80 &$-$1.07 &1.35&$b-y$ &    0.30&     0.23&     0.22&     0.22\\
HD\,195633                         &VLT/UVES                  &5894 &3.89 &$-$0.59 &1.45&$b-y$ & $-$0.29&  $-$0.05&  $-$0.06&  $-$0.06\\
HD\,205650                         &VLT/UVES                  &5714 &4.39 &$-$1.10 &0.87&$b-y$ &    0.04&     0.08&  $-$0.05&     0.02\\
HD\,298986                         &VLT/UVES                  &6086 &4.03 &$-$1.33 &1.30&$b-y$ &    0.22&     0.33& $<$ 0.33&     0.33\\
\hline                                                                          
G\,21-22                           &Subaru/HDS                &5657 &4.46 &$-$0.98 &1.35&$V-K$ & $-$0.10&  $-$0.14&     0.10&  $-$0.02\\
HD\,6268                           &Subaru/HDS                &4726 &1.14 &$-$2.63 &2.05&$b-y$ &    0.34&     0.57& $<$ 0.64&     0.57\\
HD\,94028                          &Subaru/HDS                &5926 &4.23 &$-$1.54 &1.50&$V-K$ &    0.24&     0.29&     0.28&     0.28\\
HD\,110184                         &Subaru/HDS                &4275 &0.56 &$-$2.52 &2.00&$b-y$ &    0.23& $<$ 0.45& $<$ 0.66& $<$ 0.45\\
HD\,198390                         &Subaru/HDS                &6339 &4.20 &$-$0.31 &1.92&$b-y$\tablefootmark{1}&    0.00&     0.26&     0.16&     0.21\\
HD\,201891                         &Subaru/HDS                &5827 &4.43 &$-$1.04 &1.55&$b-y$\tablefootmark{1}&    0.05&     0.04&  $-$0.01&     0.02\\
HD\,220242                         &Subaru/HDS                &6804 &4.00 &   0.02 &2.50&$b-y$ &$<$ 0.08&     0.22& $<$ 0.29&     0.22\\
\hline                                                           
BD\,+04 4551\tablefootmark{b}      &Keck/HIRES                &5990 &3.85 &$-$1.43 &1.41&$V-K$ &    0.58&     0.91&     0.95&     0.93\\
BD\,+07 4841                       &Keck/HIRES                &6187 &3.93 &$-$1.50 &1.65&$V-K$ &$<$ 0.79&     0.67&     0.68&     0.68\\
BD\,+17 4708\tablefootmark{a}      &Keck/HIRES                &5938 &3.94 &$-$1.60 &1.30&$V-K$ &    0.33&     0.48& $\cdots$&     0.48\\
BD\,+19 1185                       &Keck/HIRES                &5507 &4.41 &$-$1.02 &1.00&$b-y$ &    0.05&  $-$0.12&  $-$0.20&  $-$0.16\\
BD\,+21 607\tablefootmark{a}       &Keck/HIRES                &6100 &4.10 &$-$1.64 &1.45&$b-y$ &    0.35&     0.25& $<$ 0.29&     0.25\\
BD\,+22 396                        &Keck/HIRES                &5571 &4.26 &$-$1.12 &0.80&$V-K$ &    0.05&     0.00&     0.15&  $-$0.08\\
BD\,+37 1458                       &Keck/HIRES                &5414 &3.42 &$-$1.99 &1.10&$V-K$ &    0.28&     0.17&     0.17&     0.17\\
BD\,+51 1696                       &Keck/HIRES                &5567 &4.41 &$-$1.29 &0.65&$V-K$ &    0.28& $<$ 0.23&     0.22&     0.22\\
BD\,$-$01 306                      &Keck/HIRES                &5646 &4.32 &$-$0.90 &1.20&$V-K$ &    0.10&     0.04&  $-$0.04&     0.00\\
BD\,$-$17 484                      &Keck/HIRES                &6125 &4.06 &$-$1.54 &1.20&$V-K$ &$<$ 0.48&     0.36& $<$ 0.38&     0.36\\
G\,10-4                            &Keck/HIRES                &4974 &4.50 &$-$2.48 &1.49&$V-K$ &    0.32& $<$ 0.36&     0.25&     0.25\\
G\,24-25                           &Keck/HIRES                &5505 &3.69 &$-$1.61 &1.35&$V-K$ &    0.30&     0.24& $<$ 0.44&     0.24\\
G\,63-46                           &Keck/HIRES                &5696 &3.97 &$-$0.85 &1.00&$V-K$ &    0.12&     0.14&     0.04&     0.09\\
G\,74-5                            &Keck/HIRES                &5669 &4.32 &$-$0.87 &0.53&$V-K$ & $-$0.03&     0.02&  $-$0.10&  $-$0.04\\
G\,88-10\tablefootmark{b}          &Keck/HIRES                &5877 &4.00 &$-$2.63 &1.60&$V-K$ &$<$ 1.50& $\cdots$&     1.74&     1.74\\
G\,113-22                          &Keck/HIRES                &5565 &3.95 &$-$1.03 &1.10&$V-K$ &    0.18&     0.06&     0.04&     0.05\\
G\,126-36                          &Keck/HIRES                &5500 &4.50 &$-$1.06 &0.80&$V-K$ &    0.40&     0.27&     0.12&     0.20\\
G\,130-65                          &Keck/HIRES                &6031 &3.65 &$-$2.22 &1.41&$V-K$ &    0.45& $\cdots$& $\cdots$& $\cdots$\\
G\,153-21                          &Keck/HIRES                &5435 &4.44 &$-$0.66 &1.61&$V-K$ & $-$0.25&  $-$0.14&  $-$0.35&  $-$0.24\\
G\,180-24                          &Keck/HIRES                &5959 &4.12 &$-$1.42 &1.20&$V-K$ &    0.23&     0.43& $\cdots$&     0.43\\
G\,188-22                          &Keck/HIRES                &5896 &4.22 &$-$1.30 &1.11&$V-K$ &    0.24& $\cdots$& $<$ 0.18& $<$ 0.18\\
G\,191-55                          &Keck/HIRES                &5570 &4.11 &$-$1.81 &0.60&$V-K$ &    0.15&     0.11&     0.11&     0.11\\
G\,192-43                          &Keck/HIRES                &6181 &3.83 &$-$1.49 &1.32&$V-K$ &    0.39&     0.62&     0.62&     0.62\\
HD\,14877                          &Keck/HIRES                &5971 &4.03 &$-$0.42 &1.57&$b-y$\tablefootmark{3}& $-$0.08&  $-$0.08&  $-$0.08&  $-$0.08\\
HD\,22521                          &Keck/HIRES                &5783 &3.96 &$-$0.25 &1.55&$b-y$\tablefootmark{3}& $-$0.25&  $-$0.15&  $-$0.15&  $-$0.15\\
HD\,24289                          &Keck/HIRES                &5682 &3.48 &$-$2.08 &1.20&$b-y$ &$<$ 0.25& $<$ 0.29& $<$ 0.34& $<$ 0.29\\
HD\,24421                          &Keck/HIRES                &5987 &4.14 &$-$0.38 &1.51&$b-y$\tablefootmark{3}& $-$0.18&  $-$0.07&  $-$0.07&  $-$0.07\\
HD\,25173                          &Keck/HIRES                &5867 &4.07 &$-$0.62 &1.79&$b-y$\tablefootmark{1}&    0.02& $<$ 0.00&     0.00&     0.00\\
HD\,26421                          &Keck/HIRES                &5737 &3.98 &$-$0.39 &1.53&$b-y$\tablefootmark{3}& $-$0.10&  $-$0.11&  $-$0.19&  $-$0.15\\
HD\,28620                          &Keck/HIRES                &6101 &4.08 &$-$0.52 &1.58&$b-y$\tablefootmark{2}& $-$0.15&     0.04&     0.04&     0.04\\
HD\,30743                          &Keck/HIRES                &6294 &3.99 &$-$0.55 &1.56&$b-y$ &    0.07&     0.24&     0.22&     0.23\\
HD\,31128                          &Keck/HIRES                &5857 &4.28 &$-$1.46 &0.89&$b-y$ &    0.26& $<$ 0.31&     0.19&     0.19\\
HD\,33632                          &Keck/HIRES                &5962 &4.30 &$-$0.23 &1.56&$b-y$\tablefootmark{1}& $-$0.20&  $-$0.16&  $-$0.24&  $-$0.20\\
HD\,54717                          &Keck/HIRES                &6350 &4.26 &$-$0.44 &2.00&$b-y$\tablefootmark{1}& $-$0.05&     0.11&     0.11&     0.11\\
HD\,63333                          &Keck/HIRES                &6054 &4.25 &$-$0.38 &1.46&$b-y$\tablefootmark{3}& $-$0.11&  $-$0.02&  $-$0.05&  $-$0.04\\
HD\,68284                          &Keck/HIRES                &5832 &3.91 &$-$0.56 &1.60&$b-y$\tablefootmark{2}& $-$0.10&  $-$0.04&  $-$0.06&  $-$0.05\\
HD\,80218                          &Keck/HIRES                &6091 &4.19 &$-$0.28 &1.51&$b-y$\tablefootmark{3}& $-$0.08&  $-$0.04&     0.00&  $-$0.02\\
HD\,89125                          &Keck/HIRES                &6038 &4.25 &$-$0.36 &1.66&$b-y$\tablefootmark{1}& $-$0.19&  $-$0.12&  $-$0.12&  $-$0.12\\
HD\,91638                          &Keck/HIRES                &6160 &4.29 &$-$0.25 &1.47&$b-y$\tablefootmark{3}& $-$0.16&  $-$0.03&  $-$0.03&  $-$0.03\\
HD\,91889                          &Keck/HIRES                &6020 &4.15 &$-$0.24 &1.66&$b-y$\tablefootmark{1}& $-$0.13&  $-$0.05&  $-$0.05&  $-$0.05\\
HD\,94835                          &Keck/HIRES                &5814 &4.43 &   0.05 &1.26&$b-y$\tablefootmark{3}& $-$0.23&  $-$0.24&  $-$0.24&  $-$0.24\\
HD\,100180                         &Keck/HIRES                &5866 &4.12 &$-$0.11 &1.87&$b-y$\tablefootmark{1}& $-$0.19&  $-$0.21&  $-$0.21&  $-$0.21\\
HD\,104056                         &Keck/HIRES                &5786 &4.23 &$-$0.55 &1.30&$b-y$ &    0.13&     0.19&     0.19&     0.19\\
HD\,109303                         &Keck/HIRES                &5905 &4.10 &$-$0.47 &1.50&$b-y$\tablefootmark{1}& $-$0.19&  $-$0.07&  $-$0.08&  $-$0.08\\
HD\,118244                         &Keck/HIRES                &6234 &4.13 &$-$0.53 &1.92&$b-y$\tablefootmark{1}& $-$0.12&     0.05&     0.05&     0.05\\
HD\,134439                         &Keck/HIRES                &5029 &4.87 &$-$1.28 &0.60&$b-y$ & $-$0.18&  $-$0.14&  $-$0.14&  $-$0.14\\
HD\,134440                         &Keck/HIRES                &4851 &4.99 &$-$1.32 &1.20&$b-y$ & $-$0.06& $<$ 0.09&  $-$0.06&  $-$0.06\\
HD\,186379                         &Keck/HIRES                &5806 &3.99 &$-$0.39 &1.54&$b-y$\tablefootmark{3}& $-$0.21&  $-$0.12&  $-$0.18&  $-$0.15\\
HD\,194598                         &Keck/HIRES                &5943 &4.12 &$-$1.23 &1.50&$b-y$ &    0.22&     0.30&     0.30&     0.30\\
HD\,200580                         &Keck/HIRES                &5829 &4.39 &$-$0.54 &1.72&$b-y$\tablefootmark{1}& $-$0.18&  $-$0.13&  $-$0.18&  $-$0.16\\
HD\,202884                         &Keck/HIRES                &6141 &4.36 &$-$0.24 &1.42&$b-y$\tablefootmark{3}& $-$0.08&  $-$0.11&  $-$0.11&  $-$0.11\\
HD\,204712                         &Keck/HIRES                &5888 &4.12 &$-$0.48 &1.49&$b-y$\tablefootmark{3}& $-$0.09&  $-$0.03&  $-$0.06&  $-$0.04\\
HD\,209320                         &Keck/HIRES                &5994 &4.14 &$-$0.18 &1.51&$b-y$\tablefootmark{3}& $-$0.24&  $-$0.01&  $-$0.05&  $-$0.03\\
HD\,209858                         &Keck/HIRES                &5911 &4.26 &$-$0.27 &1.40&$b-y$\tablefootmark{3}& $-$0.18&  $-$0.20&  $-$0.20&  $-$0.20\\
HD\,215442                         &Keck/HIRES                &5872 &3.80 &$-$0.22 &1.69&$b-y$\tablefootmark{3}& $-$0.32&  $-$0.21&  $-$0.21&  $-$0.21\\
HD\,241253                         &Keck/HIRES                &5877 &4.08 &$-$1.12 &1.02&$V-K$ &    0.22&     0.25&     0.23&     0.24\\
HD\,247297                         &Keck/HIRES                &5449 &3.45 &$-$0.56 &1.15&$b-y$ & $-$0.24&  $-$0.14&  $-$0.18&  $-$0.16\\
HD\,345957                         &Keck/HIRES                &5752 &3.91 &$-$1.30 &1.04&$b-y$ &    0.15&     0.06&     0.06&     0.06\\
Ross\,390                          &Keck/HIRES                &5299 &4.36 &$-$1.19 &0.95&$V-K$ &    0.15&     0.16& $\cdots$&     0.16\\
Ross\,797                          &Keck/HIRES                &6255 &3.87 &$-$1.25 &0.80&$V-K$ &    0.80&     0.80&     0.84&     0.82\\
    \end{longtable}
    \tablefoot{
    $b-y$ and $V-K$ in the column of method indicate $T_{\rm eff}$ derived from $b-y$ and $V-K$, respectively.\\
    \tablefoottext{1}{Stellar parameters taken from \citet{Chen2000}.}
    \tablefoottext{2}{Stellar parameters taken from \citet{Chen2001}.}
    \tablefoottext{3}{Stellar parameters taken from \citet{Reddy2003}.}
        \tablefoottext{a}{Stars with lower spectral S/N and therefore the abundances are more uncertain than the average.}
        \tablefoottext{b}{Stars in known binary systems.}
        \tablefoottext{c}{Spectra taken from the UVES-POP survey.}
    }
}
}

\section{Abundances}

The ultraviolet \ion{Ag}{i} $\lambda$ 3280, 3382 \AA,\ and \ion{Pd}{i}
    $\lambda$ 3404\,\AA\ regions are heavily blended by atomic and molecular
    lines.
We used the spectral synthesis method with the IDL/Fortran {\sc SIU} software package
    \citep{Reetz1991} to derive the Pd and Ag abundances of the sample stars.
The relevant atomic data of \ion{Pd}{i} and \ion{Ag}{i} used in this work were
    presented in Table \ref{tbl-atom}.
For silver, the $\log gf$ values were taken from \cite{Ross1972}, where the
    hyperfine structure of both lines were taken into account.
The $\log gf$ value of palladium was taken from the VALD\footnote{Vienna Atomic
    Lines Database, available at
    http://vald.astro.univie.ac.at/vald/php/vald.php} database.

The overall $\log{gf}$ values for \ion{Ag}{i} $\lambda$ 3280.68 \AA\ and
    $\lambda$ 3382.90\,\AA\ are 0.005 and 0.002\,dex higher than those of
    \citet{Hansen2012}, respectively.
While for \ion{Pd}{i}, our adopted $\log{gf}$ is 0.02 dex lower.
The van der Waals damping constants $\log C_6$ of both silver and palladium
    lines were calculated according to the \citet{Anstee1991,Anstee1995}
    interpolation tables.
The blended atomic and molecular lines involve NH, \ion{Fe}{i}, \ion{Fe}{ii},
    \ion{Ni}{i}, \ion{Ni}{ii}, \ion{V}{i}, \ion{V}{ii}, \ion{Cr}{i},
    \ion{Ti}{i}, and \ion{Ti}{ii}.
We have considered the influence induced by the \ion{Zr}{ii} line
    on the red wing of \ion{Ag}{i} $\lambda 3280$\,\AA\ by synthesizing the
    spectra with oscillation strength $\log{gf}=-1.1$, $-1.5$ (same as
    \citealt{Hansen2012}) and without this transition.
We found that the differences are less than 0.03\,dex for the deduced Ag
    abundance, and we did not include \ion{Zr}{ii} in our line list.
The line information was taken from Kurucz database with slight adjustments.
To minimize the systematic offsets and make a direct comparison with
    \citet{Hansen2012}, we adopted the same solar abundance values
    ($\log\epsilon({\rm Ag})_\odot=0.94$; $\log\epsilon({\rm Pd})_\odot=1.57$)
    in our analysis as those used in their study.

\begin{table}[htbp]
\caption{Atomic data}
\label{tbl-atom}
\begin{tabular}{ccccc}
\hline
\hline
      & $\lambda $       & $E_\mathrm{low}$  & $\log gf$  & $\log C_6$ \\
      & $(\mathrm{\AA})$ &              (eV) &            &            \\
\hline
Ag I  & 3280.682 &  0.00 & -0.452 &  -31.795   \\
Ag I  & 3280.683 &  0.00 & -0.432 &  -31.795   \\
Ag I  & 3280.688 &  0.00 & -0.912 &  -31.795   \\
Ag I  & 3280.690 &  0.00 & -0.933 &  -31.795   \\
\hline
Ag I  & 3382.899 &  0.00 & -0.755 &  -31.829   \\
Ag I  & 3382.900 &  0.00 & -0.754 &  -31.829   \\
Ag I  & 3382.905 &  0.00 & -1.235 &  -31.829   \\
Ag I  & 3382.908 &  0.00 & -1.256 &  -31.829   \\
\hline
Pd I  & 3404.582 &  0.81 &  0.300 &  -31.822   \\
\hline
\end{tabular}
\end{table}

\begin{figure*}[htbp]\centering
\includegraphics[width=18cm]{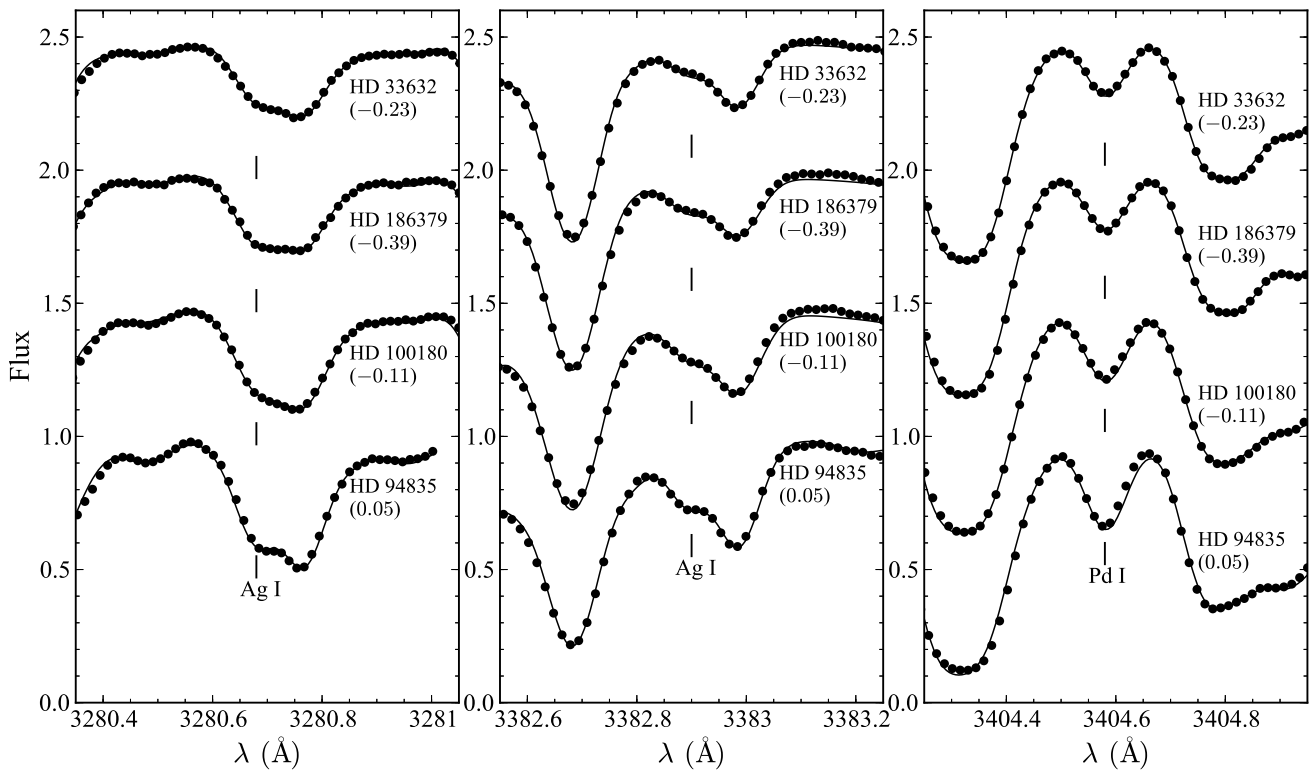}
\caption{
    Spectral synthesis for four thin disk stars in our sample.
    From top to bottom: HD\,33632, HD\,186379, HD\,100180, and HD\,94835.
    Solid dots represent the observed spectra with Keck/HIRES, and the solid
        lines are the synthesis spectra.
    Vertical lines indicate the location of \ion{Ag}{i} and \ion{Pd}{i} lines.
    Offsets of 0.5 were added from the bottom to the top to improve the
        visualization.
}
\label{fig-4stars}
\end{figure*}

\onlfig{
\begin{figure*}
\includegraphics[width=18cm]{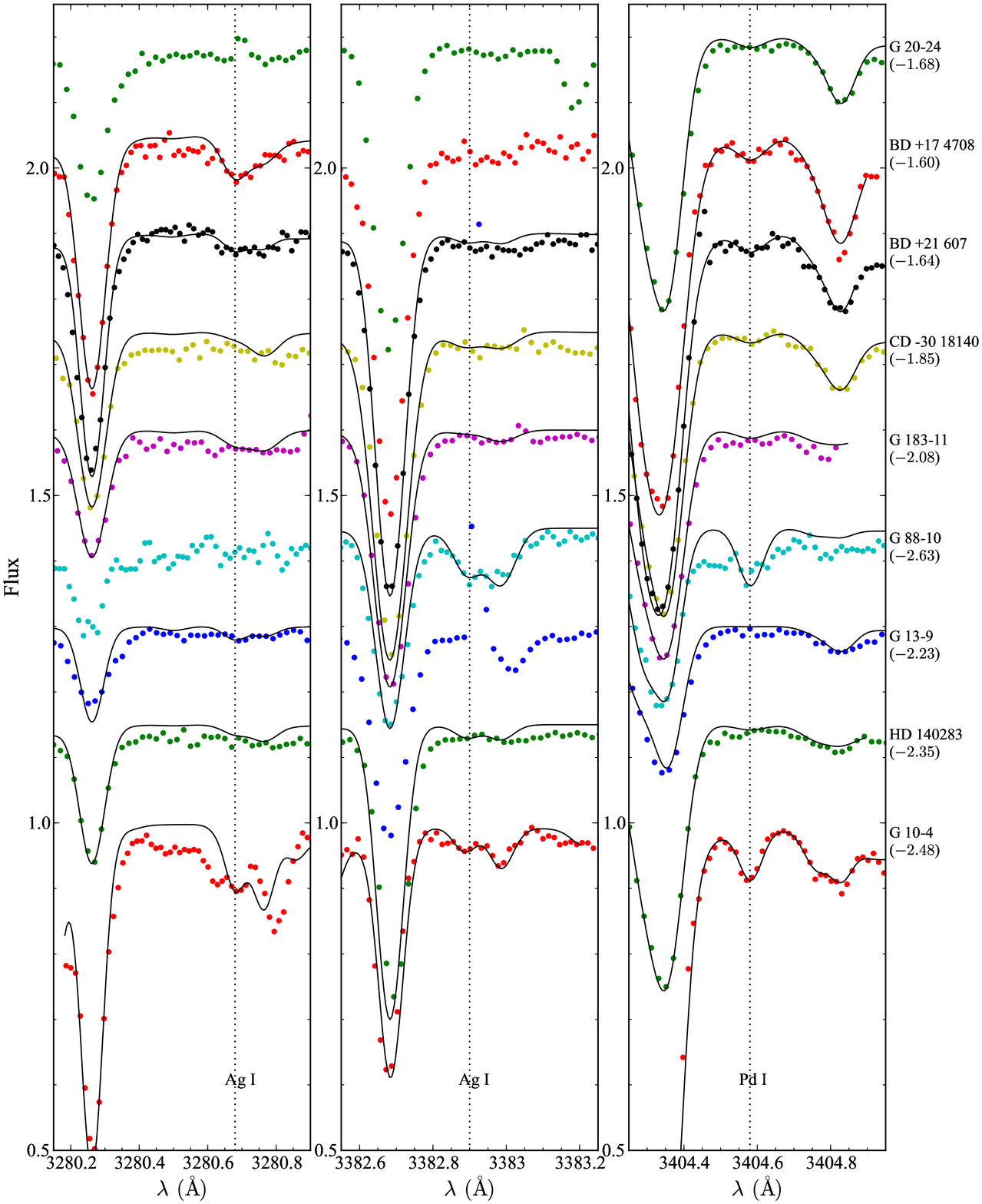}
\caption{
    Examples of spectral synthesis of metal-poor stars with [Fe/H] $<-1.5$.
    Colour dots and solid lines indicate observed and synthesized spectra,
    respectively.
    Vertical dash lines are the locations of Pd and Ag lines analyzed in
    this work.
    The \ion{Ag}{i} $\lambda\,3382.9$\,\AA\ region of G\,88-10 and the
    \ion{Ag}{i} $\lambda\,3382.9$\,\AA\ regions of G\,20-24 and BD\,+21\,607
    are severely affected by noise or do not show any absorption feature.
    The pixels in the \ion{Ag}{i} $\lambda\,3382.9$\,\AA\ region of
    G\,13-9 and \ion{Ag}{i} $\lambda\,3280.7$\,\AA\ region of G\,20-24 and
    BD\,+17\,4708 were affected by cosmic rays, preventing us from any spectral
    synthesis.
    Ag abundance detections were obtained only for BD\,+17\,4708 and
    BD\,+21\,607 with \ion{Ag}{i} $\lambda\,3280.7$\,\AA\  and only for
    G\,88-10 and G\,10-4 with \ion{Ag}{i} $\lambda\,3382.9$\,\AA\ in this
    figure.
    Pd abundance detection were obtained for four stars with \ion{Pd}{i}
    $\lambda\,3404.6$\,\AA\ (see online Table \ref{tab-abundance}).
}
\label{fig-9stars}
\end{figure*}
}

\begin{figure}[htbp]\centering
\includegraphics[width=9cm]{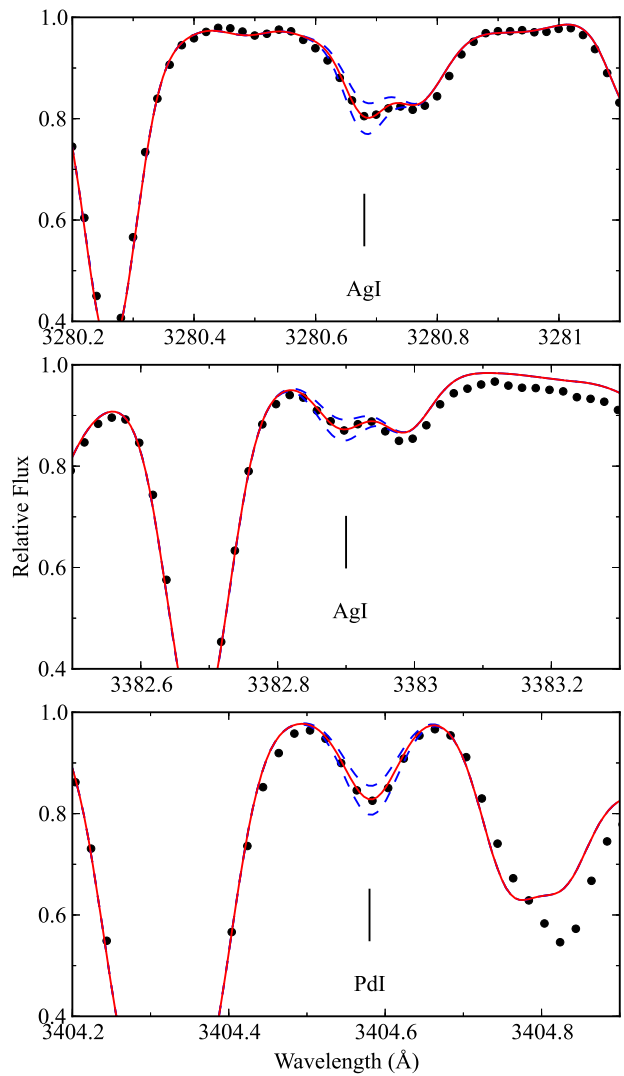}
\caption{
    Example of spectral synthesis around Ag and Pd regions used in this
    analysis.
    The dotted lines and red solid lines represent the observed spectra and
    synthesis spectra of HD\,76932 ([Fe/H] = $-0.96$), respectively.
    The blue solid lines represent the synthesis spectra with [X/Fe] $\pm$ 0.1
    dex.
}
\label{fig-syn}
\end{figure}

\begin{figure*}[htbp]\centering
\includegraphics[width=18cm]{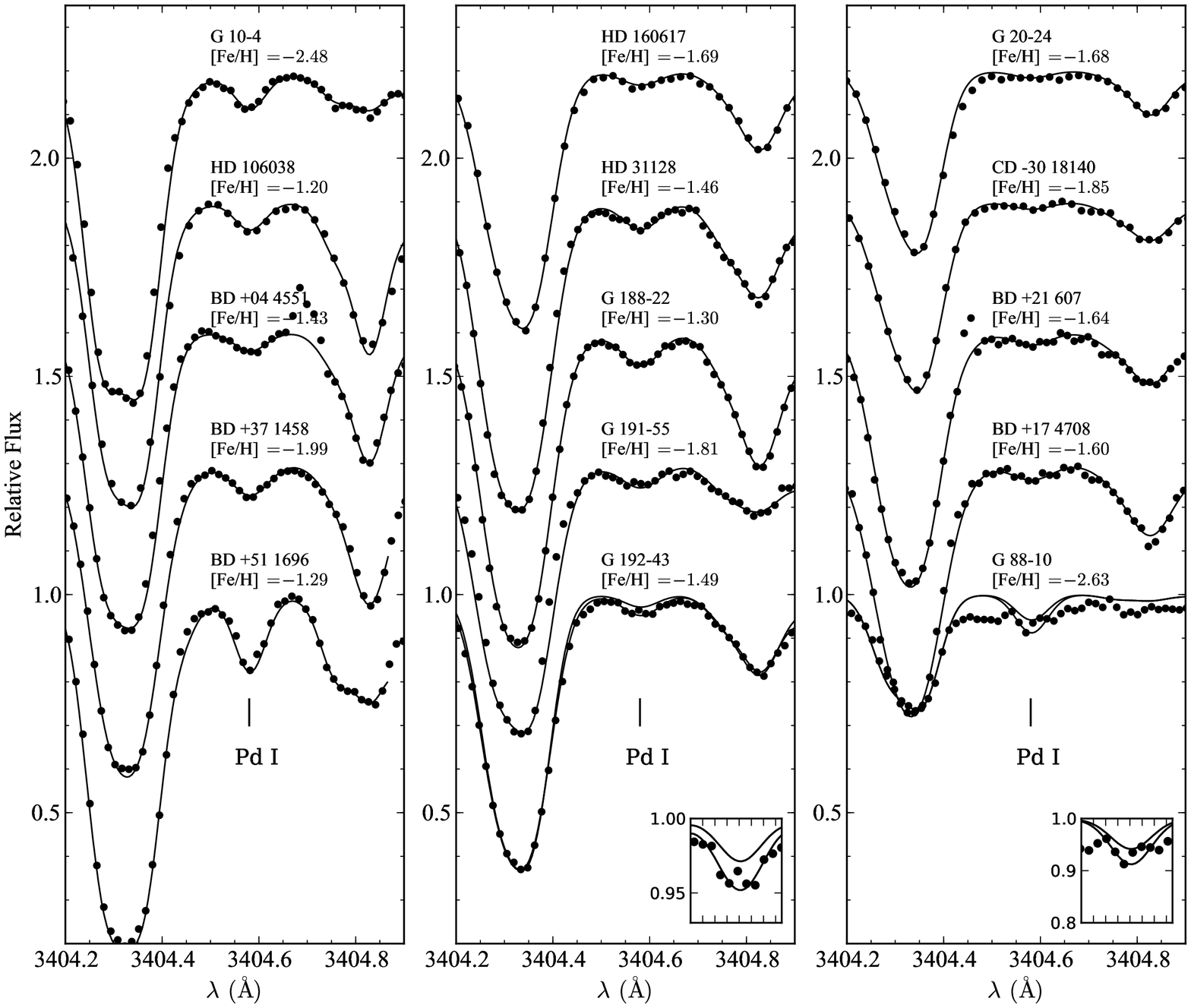}
\caption{
    Examples of spectral synthesis around \ion{Pd}{i} $\lambda 3404.9 \AA$ used
    in this analysis.
    The dots represent the observational data and solid lines the
    synthesis spectra.
    Offsets of 0.3 were added for better visualization.
    See also the online figure for details.
    For G\,192-43, the absorption feature exhibits a W shape, as shown in the
    partial enlarged drawing of the spectrum in the lower right of the middle
    panel.
    Two synthesis spectra with [Pd/Fe] = 0.20 and 0.39 are overplotted,
    respectively.
    A similar plot was shown for G\,88-10 (right panel), for which the lower
    synthesis ([Pd/Fe] = 1.50) was adopted as an upper limit.
}
\label{fig-syn-pd1}
\end{figure*}

Figures \ref{fig-4stars}, \ref{fig-9stars}, \ref{fig-syn} and \ref{fig-syn-pd1}
    show the examples of spectral synthesis of the \ion{Ag}{i} and \ion{Pd}{i}
    lines for our sample stars.
All of the four stars in Fig. \ref{fig-4stars} have solar metallcities or are
    mildly metal-poor, with [Fe/H] ranging from $-0.39$ to +0.05.
According to the kinetic criteria in section \ref{kinetics}, they belong to the
    thin disk.
Although the \ion{Ag}{i} lines were heavily blended with \ion{Mn}{i} ($\lambda$
    3280.77 \AA) or \ion{Fe}{i} ($\lambda$ 3382.98 \AA), the spectral region
    can be well fitted by assuming Gaussian instrumental profiles.

\subsection{Error estimation}
\label{error}

The errors in the derived chemical abundances are mainly due to the
    uncertainties in the stellar parameters and the continuum placement.
Considering that the S/Ns of most spectra are higher than 100, the estimated
    abundance uncertainties caused by continuum location are around 0.15\,dex.
The errors caused by the uncertainties in the stellar atmospheric parameters are
    about 0.12\,dex for Pd and 0.13\,dex for Ag.
Table \ref{tbl-err} gives the abundance differences due to deviations of the
    effective temperature of 80\,K, the surface gravity of 0.15\,dex, the iron
    abundance of 0.10\,dex and the microturbulent velocity of 0.2\,km\,s$^{-1}$
    for a typical star, HD\,76932.
By quadratic summing of the above uncertainties, the errors of Pd and Ag
    abundances are estimated to be 0.19 and 0.20\,dex, respectively.

\begin{table*}[htbp]
    \caption{Abundance uncertainties of HD\,76932, with $T_{\rm eff}=5849$\,K,
    $\log{g}=4.05$, [Fe/H] = $-0.96$, and $\xi$ = 1.6\,km\,s$^{-1}$.}
\label{tbl-err}
\centering
\begin{tabular}{lrrrrr}
\hline
\hline
  Element ratios                          &
  $\Delta T_{\rm eff}$ ($\pm 80$ K)       &
  $\Delta\log g$ ($\pm 0.15$)             &
  $\Delta$[Fe/H] ($\pm 0.10$)             &
  $\Delta{\xi}$ ($\pm 0.2$ km\,s$^{-1}$)  &
  $\sigma_{\rm Total}$                   \\
\hline
 [Ag/H]$_{3382.9}$  & $\pm0.10$  & $\mp0.01$ &  $\pm0.06$  &  $\pm0.01$  & $\pm0.11$ \\

 [Ag/H]$_{3280.7}$  & $\pm0.10$  & $\pm0.00$ &  $\pm0.08$  &  $\pm0.01$  & $\pm0.13$ \\

 [Pd/H]$_{3404.6}$  & $\pm0.10$  & $\pm0.01$ &  $\pm0.07$  &  $\pm0.01$  & $\pm0.12$ \\
\hline
\end{tabular}
\end{table*}

\subsection{Comparison with the literature}

\begin{figure}[htbp]\centering
\includegraphics[width=4.3cm]{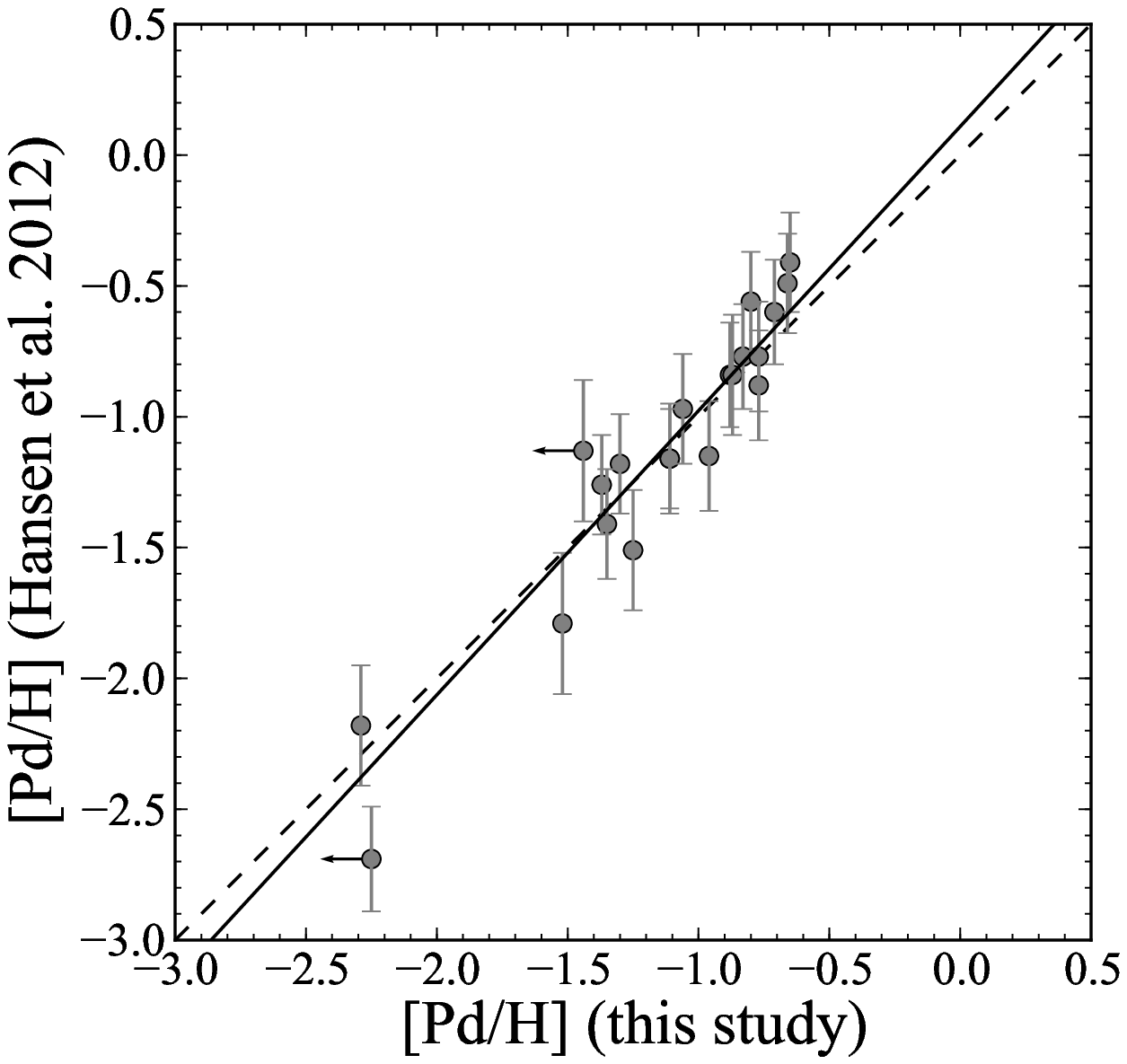}
\includegraphics[width=4.3cm]{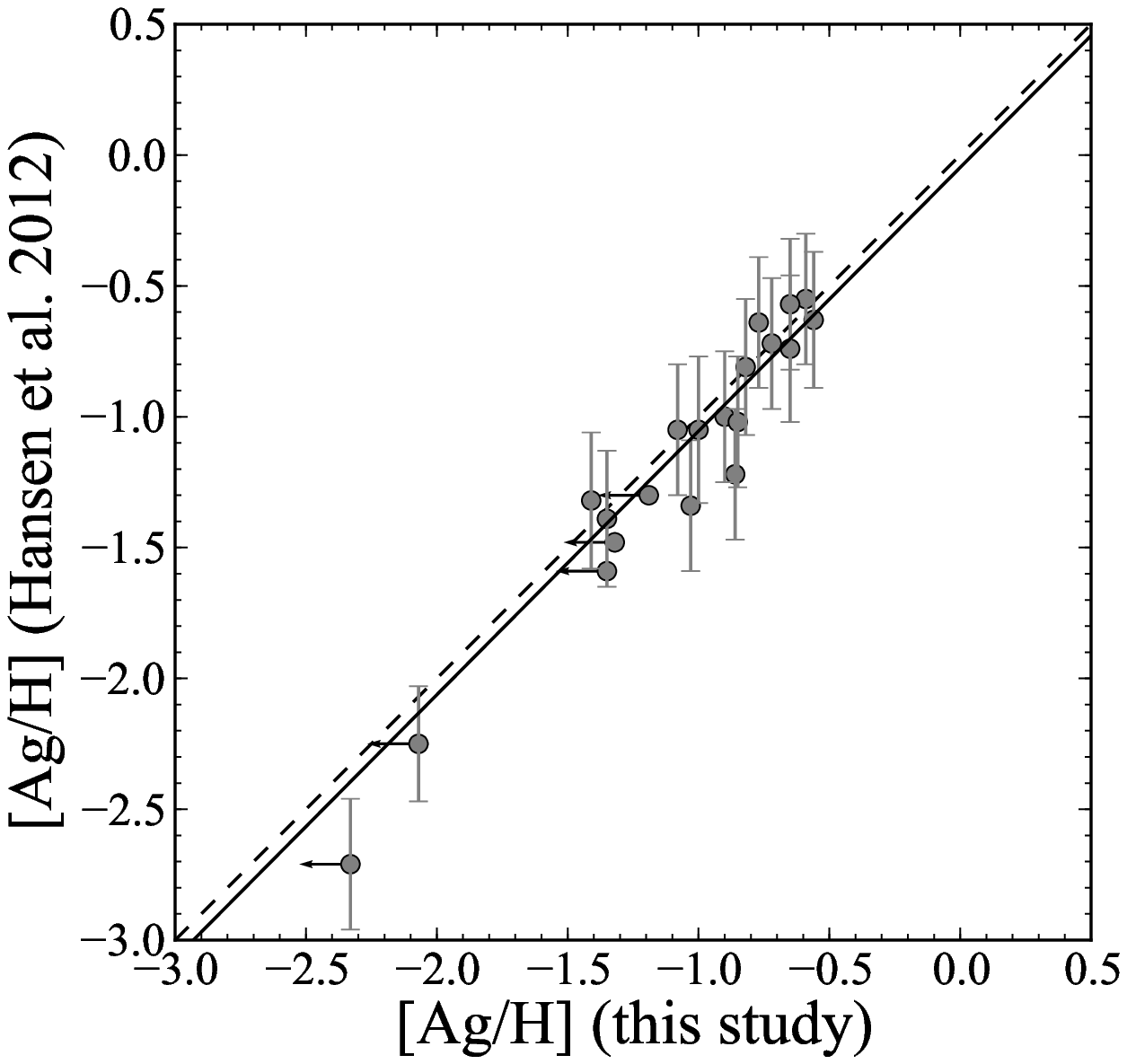}
\caption{
    Comparison of Ag and Pd abundances of common stars in this study with those
    in \cite{Hansen2012}.
    The dashed lines represent 1:1 ratios, and solid lines represent the linear
    least-square fitting functions
    $y = (1.09\pm0.09)x + (0.11\pm0.10)$ for [Pd/H], and
    $y = (1.01\pm0.15)x + (-0.05\pm0.14)$ for [Ag/H], where $x$ denotes our
    abundances and $y$ denotes those of \citet{Hansen2012}.
}
\label{fig-comp-abun}
\end{figure}

\cite{Hansen2011} and the follow-up study \citep{Hansen2012} derived Pd and Ag
    abundances for 34 dwarfs and 23 giants, of which 23 dwarfs and 2 giants are
    also included in our sample.
We compared our abundances with the results of \cite{Hansen2012} for common
    stars in Fig. \ref{fig-comp-abun}.
We found excellent agreement with the \citet{Hansen2012} results.
The mean differences are
    $\left<\Delta{\rm [Pd/H]}\right>=+0.017\pm0.145$\,dex and
    $\left<\Delta{\rm [Ag/H]}\right>=-0.054\pm0.135$\,dex,
    and both of the scatters are within the errors.

For G\,183-11 and HD\,140283, \citet{Hansen2012} did not report any Ag abundance,
    while we gave the upper limits based on the shallow absorption features
    exhibit in the spectra.
The poor S/Ns and difficulties with locating the continuum placement properly has
    prevented any accurate measurements on these lines.
While for CD\,$-$30\,18140, we gave \ion{Pd}{i} abundance value based on
    a weak line, because the absorption feature has exceeded the 3$\sigma$ level
    below the continuum so cannot be fully explained by spectral noise.
Similar features were also found for BD\,+21\,607 and BD\,+17\,4708.
For G\,20-24, we derived an upper limit of \ion{Pd}{i}, and yet it is inconsistent
    with \citet{Hansen2012} ([Pd/Fe] = $0.77\pm0.27$).
We attribute this conflict to the large differences in the parameters determined
    by \cite{Hansen2012} ($T_{\rm eff}=6482$\,K, [Fe/H] = $-1.90$\,dex) and this
    work (6141\,K, $-1.68$\,dex).
Their spectral synthesis are plotted in the righthand panel of Fig.
    \ref{fig-syn-pd1}.

In addition, for HD\,103723, HD\,106038, and HD\,113679, our \ion{Pd}{i}
    abundances are below the 1$\sigma$ lower limits of \citet{Hansen2012}.
This is mainly due to the relative large differences in $T_{\rm eff}$ and/or the
    deviations in [Fe/H], as discussed in Sect. \ref{sect-comp-para}.

Another source of deviations in the Ag abundance may arise from the two-level
    hyperfine splitting data of \cite{Ross1972} adopted by this work and the
    three levels by \cite{Hansen2012}.
We selected 12 stars that span a wide range of $T_{\rm eff}$ and metallicities,
    and derived their Ag abundances with the above two sets of hyperfine
    splitting data.
We found that the differences are within 0.01\,dex, so much smaller than the
    claimed uncertainties of Ag in any studies, so they can be neglected.

\citet{Peterson2013} also analyzed a sample of 29 turnoff stars and determined
    Pd abundances for 14 of them.
There are 16 stars shared with our sample.
For seven stars (BD\,+17\,4708, BD\,+21\,607, BD\,+37\,1458, G\,188-22,
    HD\,31128, HD\,106038, HD\,160617), they did not report Pd abundance,
    but we gave detections.
We plotted the \ion{Pd}{i} regions in figure \ref{fig-syn-pd1}.
For the remaining nine stars, our Pd abundances agree with their values very
    closely, except for G\,191-55 and G\,192-43.
For G\,191-55, \citet{Peterson2013} gave [Pd/Fe] = 0.3, while we found a lower
    value of 0.11\,dex.
The difference again results from the large deviations of stellar parameters,
    especially $T_{\rm eff}$ (6000\,K in \citealt{Peterson2013}; 5570\,K in our
    work).
For G\,192-43 ([Fe/H] $\simeq-1.4$), where \citet{Peterson2013} detected [Pd/Fe]
    = 0.2, we found [Pd/Fe] = 0.39 using the \ion{Pd}{i} line of the same
    Keck/HIRES spectra.
The line profile of \ion{Pd}{i} $\lambda3404.5\,\AA$ exhibits a W-shape, as shown
    in figure \ref{fig-syn-pd1}.
We found that the synthesis spectrum with [Pd/Fe] = 0.2 and their parameters
    ($T_{\rm eff}/\log{g}/\mathrm{[Fe/H]}/\xi=6200/3.9/-1.5/1.2$) cannot explain
    the tip of the line core fully, whereas we have adopted the lower profile
    with slightly higher abundance.

\section{Results and discussion}
\label{discussion}

\begin{figure}[htbp]\centering
\includegraphics[width=9cm]{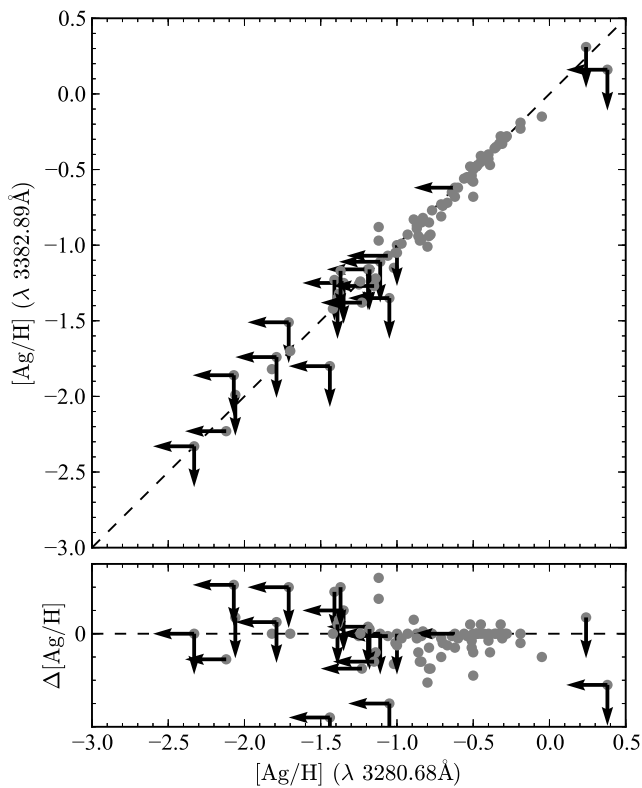}
\caption{
    Comparison of abundance values derived from \ion{Ag}{i} $\lambda 3280.68\,
    \AA$ and \ion{Ag}{i} $\lambda 3382.89\,\AA$.
    Arrows represent the upper or lower limits.
    The outlier with $x=-2.01$ and $y=-0.7$ is a spectroscopic binary G\,88-10
    (see discussions in section \ref{discussion}).
}
\label{fig-ag-comp}
\end{figure}

The resulting Pd and Ag abundances are listed in Table \ref{tab-abundance},
    which is only available online.
For a few stars, we only get the upper or lower limits of abundance values because the
    lines are either too weak or are severely blended.
The comparison of the \ion{Ag}{i} abundances derived with the lines of $\lambda$
    3280 \AA\ and $\lambda$ 3382 \AA\ is shown in figure \ref{fig-ag-comp}.
The mean difference $\left<{\rm [Ag/H]}_{3382} - {\rm [Ag/H]}_{3280}\right>$ is
    $-0.024\pm0.068$ dex, and there is no significant trend in the residual.

\begin{figure}[htbp]
\includegraphics[width=9cm]{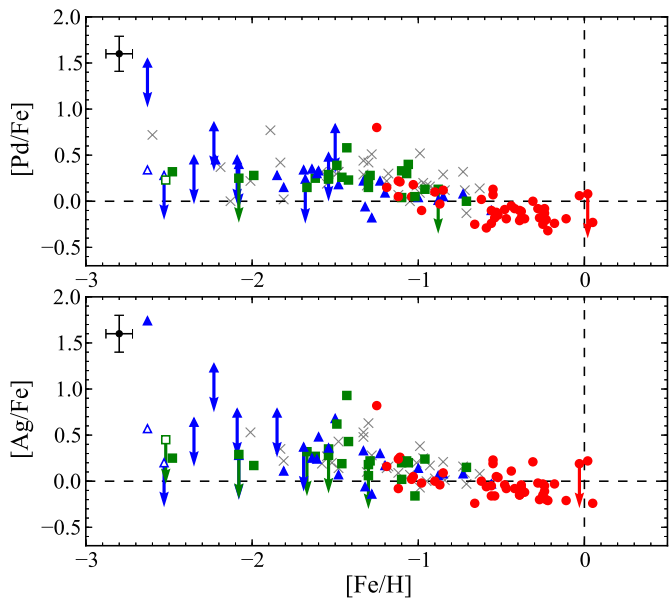}
\caption{
    [Pd/Fe] (upper panel) and [Ag/Fe] (lower panel) v.s. [Fe/H].
    Red, green, and blue points represent thin disk, thick disk, and halo stars.
    Solid points denote dwarfs, while open circles denote the three giants in
    our sample.
    The arrows represent the lower or upper limits of the data points.
    The typical errors (see section \ref{error}) are shown in the upper left
    corner of each panel.
    The dwarfs in \citet{Hansen2012} are also plotted as crosses in this figure.
}
\label{fig-abun}
\end{figure}

The abundance results of [Pd/Fe] and [Ag/Fe] versus [Fe/H] are plotted in figure
    \ref{fig-abun}.
The [Ag/Fe] abundances are the averages of the values derived from the two
    \ion{Ag}{i} lines, if neither of them are upper or lower limits.
Otherwise only the values without limits are adopted.
In general, [Pd/Fe] and [Ag/Fe] exihibt very similar behaviors with [Fe/H].
Both of them show flat trends above [Fe/H] $\gtrsim-0.6$, until solar
    metallicities up to [Fe/H] $\simeq$+0.1, where the sample is dominated by
    thin disk stars.
While the thick disk and halo stars are well mixed at [Fe/H] $<-1.0$, and
    both of the abundance ratios slowly increase with decreasing [Fe/H].
The large star-to-star scatters (0.16\,dex for [Pd/Fe] and 0.30\,dex for [Ag/Fe])
    at [Fe/H] $<-1.0$ are not likely due to our internal errors but are probably
    caused by the inhomogeneous mixing of newly produced nuclides in the early
    Galaxy.
And the descents of [Pd/Fe] and [Ag/Fe] with [Fe/H] at [Fe/H] $\simeq-0.6$ may
be    attributed to the rise of Fe in the SNIa yield.

\begin{figure}[htbp]
\includegraphics[width=9cm]{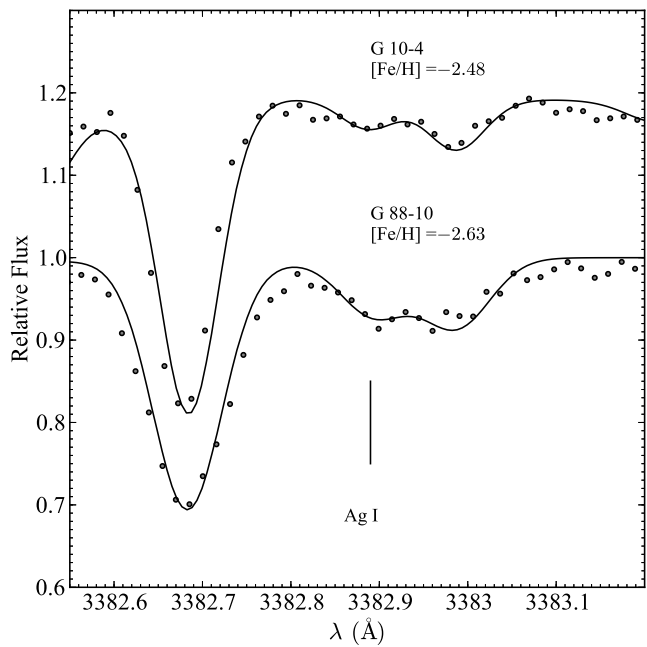}
\caption{
    Spectral synthesis of \ion{Ag}{i} region for two stars, G\,88-10 and G\,10-4,
        at the metal-poor end in our sample.
    Both of them have metallicities of [Fe/H] $\simeq-2.5$, and G\,88-10 is a
        spectroscopic binary (see text).
}
\label{fig-syn-ag}
\end{figure}

We report the detections of Ag for a few metal-poor dwarfs with [Fe/H]
    $\lesssim-2.5$, although the number is small.
For example, G\,88-10 ([Fe/H] = $-2.63$) has an exceptional overabundance of
    [Ag/Fe] = +1.74, and the Pd abundance is likely to be overabundant as well
    ([Pd/Fe] $<+1.50$), as shown in Figs. \ref{fig-syn-pd1} and
    \ref{fig-syn-ag}.
The enhancements of both Pd and Ag provide evidence of a good correlation between
    these two elements, even in such a very metal-poor environment.
Furthermore, G\,88-10 is listed as a close binary system with $P=20.6$ days in
    the SB9 catalog \citep{Goldberg2002}.
Nevertheless, the overabundances of Ag and Pd (fairly high upper limit) are probably not
    due to the photospheric pollution via material transfer between the
    two companions, because this process is thought to be mainly responsible for the
    enhancements of $s$-process elements rather than $r$-process.
Anyhow, the anomaly of G\,88-10 deserves further investigations by connecting Pd
    and Ag with other elements produced by different nucleosynthesis processes,
    such as Ba and Eu.
Another metal-poor star with [Fe/H] $\simeq-2.5$ is G\,10-4, for which we
    detected \ion{Pd}{i} line at $\lambda$ 3404.5 \AA\ (figure
    \ref{fig-syn-pd1}) and \ion{Ag}{i} line at $\lambda$ 3382.9 \AA\ (figure
    \ref{fig-syn-ag}) based on Keck/HIRES spectra.

\begin{figure}[htbp]
    \centering
    \vspace{-0.5cm}
    \includegraphics[width=9cm]{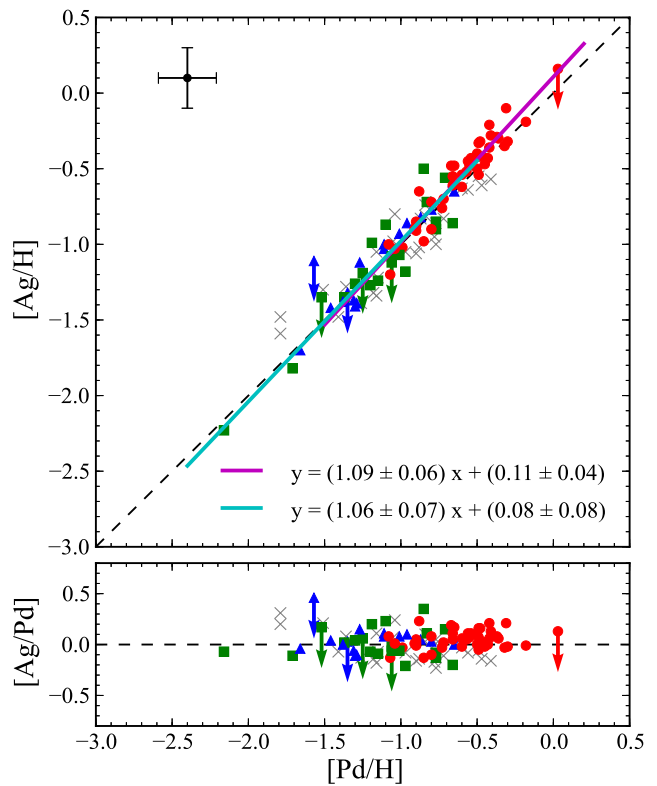}
    \caption{
        Relation of [Ag/H] versus [Pd/H] ({\it upper panel}) and [Ag/Pd]
        versus [Pd/H] ({\it lower panel}) for the dwarfs in our sample.
        {\it Red, green, and blue dots} denote the sample stars that classified
        as thin disk, thick disk, and halo components by their kinetic
        properties, respectively (see section \ref{kinetics}).
        The dashed line represents the 1:1 ratio, and the solid lines are the
        least-square linear fitting of the two subsamples of dwarfs
        ({\it magenta}: for thin disk stars; {\it cyan}: for thick disk + halo
        stars).
        The dwarfs in \citet{Hansen2012} are overplotted as {\it crosses} in this
        plot.
        The typical errors of Pd and Ag abundances are also shown in the upper
        left corner.
    }
    \label{fig-agpd}
\end{figure}

In figure \ref{fig-agpd} we plotted both the ratios of [Ag/H] versus [Pd/H] with
    colors coded by various Galactic populations.
After excluding the three giants and the dwarfs with only upper or lower limits
    of Ag or Pd abundances, we found a least-square linear fitting of [Ag/H] =
    ($1.07\pm0.03$) [Pd/H] + ($0.09\pm0.03$) for the entire sample.
This is in general consistent with the slope of 0.97 by accounting for the
    uncertainties for the sample of dwarf + giant stars found by
    \citet{Hansen2011}.
Furthermore, we found that the slope of Ag v.s. Pd stays constant at a wide
    abundance ratio range of $-2.2\lesssim\mathrm{[Pd/H]}\lesssim0.1$ in the
    Galactic dwarfs.
The mixing thick disk + halo stars almost show the same trend as that of thin
    disk stars.
If the thick disk and halo stars are seperated, the linear fits in figure
    \ref{fig-agpd} becomes
    $y=(1.05\pm0.10)x + (0.05\pm0.12)$ (thick disk) and
    $y=(1.09\pm0.07)x + (0.13\pm0.08)$ (halo), respectively.

It is well-known that the chemical evolution were dominated by the $r$-process
    in the early Galaxy.
As the Galactic metallicity increases, the contribution of the $s$-process
    becomes significant because the time scales of low-mass AGB stars are
    considered to be much longer than that of the $r$-process
    \citep[e.g.,][]{Burris2000}.
\citet{Arlandini1999} predicted that 46\% of Pd and 20\% of Ag in the solar
    system are produced by the $s$-process, and \citet{Bisterzo2011} got similar
    results (53.1\% for Pd and 22.1\% for Ag).
Both of the calculations led to a slope of [Ag/H] versus [Pd/H] less than unity
    above [Fe/H] $\simeq-1$ because the $s$-process produces roughly the same
    amount of Pd as does the $r$-process, but this fraction is only 1/5 for Ag.
However, our results do not confirm the predictions of these stellar
    nucleosynthesis models.
The trends in Ag vs. Pd in the Galactic dwarfs stay constant from a very
    metal-poor environment of $-2.6$ until the solar metallicity.

\section{Conclusions}

Based on the archive near-UV spectra obtained with Keck/HIRES, Subaru/HDS, and
    VLT/UVES, we analyzed a large sample of stars spanning the metallicity
    range of $-2.6\lesssim$ [Fe/H] $\lesssim+0.1$.
We reported the photospheric abundances of palladium (Pd) and silver (Ag) for 83
    and 79 stars, respectively, including several dwarfs with [Fe/H] $<-2.0$,
    which have seldom been detected in previous studies.
The most metal-depleted dwarf with detected Ag abundance is down to [Fe/H]
    $\simeq-2.6$.
Meanwhile, our study investigated these two elements for the first time for a
    group of stars with solar and super-solar metallicities.

Our sample has increased the number of dwarfs with known Pd and Ag abundances by
    a factor of $\sim$2.
It was found that both [Pd/Fe] and [Ag/Fe] show flat trends with metallicities
    around $-0.6\lesssim$ [Fe/H] $\lesssim+0.1$ and slowly increase with
    decreasing metallicity below [Fe/H] $\simeq-0.6$.
In metal-poor stars with [Fe/H] $<-1.5$, [Pd/Fe] and [Ag/Fe] ratios are
    enhanced by $\sim$0.3\,dex, and they show large star-to-star dispersions of
    $\sim$0.3\,dex.
On the other hand, the amount of Pd in the Galactic dwarfs grows at nearly the same
    speed as does Ag at the whole [Pd/H] ranging from $-2.2$ to $-0.2$, and
    good correlations between [Ag/H] and [Pd/H] were found for different stellar
    populations.
It seems that the trends in [Ag/H] v.s. [Pd/H] are the same within their errors
    for thin disk, thick disk, and halo stars.
These facts imply that the two elements are synthesized by similar processes
    during the Galactic chemical evolution history, and our results do not
    support the theoretical predictions that Pd and Ag have different
    contributions from the $s$-process.

\begin{acknowledgements}
This research is supported by the National Natural Science Foundation of China
    under grant Nos. 11321064, 11233004, 11473033, and U1331122.
The data used in this research were obtained from the Keck Observatory Archive
    (KOA), the Subaru-Mitaka-Okayama-Kiso Archive (SMOKA), and the ESO Science
    Archive Facility.
KOA is operated by the W. M. Keck Observatory and the NASA Exoplanet Science
    Institute (NExScI), under contract with the National Aeronautics and Space
    Administration.
SMOKA is operated by the Astronomy Data Center, National Astronomical
    Observatory of Japan.
The observations of Keck/HIRES spectra were made under program ID H41aH,
    H11aH, H03aH, H177Hb, H269Hb, and H233Hb (P.I. A.Boesgaard).
The observations of Subaru/HDS spectra were made under program ID o01319,
    o03310 (P.I. A.Boesgaard), o03429 (P.I. S.Honda), and o07136 (P.I. S.Wanajo).
The observations of VLT/UVES spectra were done under program IDs 65.L-0507 and
    67.D-0439 (P.I. F.Primas).
We thank all the above P.I.s for obtaining the data that made this study
    possible.
We thank the anonymous referee for the essential suggestions that
    improved this article.
X.S.Wu thanks Dr. Kefeng Tan for providing the reduced UVES 1-D spectra.
This research made use of the SIMBAD database, which is operated at the CDS,
    Strasbourg, France, and the data products from the Two Micron All Sky
    Survey, which is a joint project of the University of Massachusetts and the
    Infrared Processing and Analysis Center/California Institute of Technology,
    funded by the National Aeronautics and Space Administration and the National
    Science Foundation.
    
\end{acknowledgements}

\bibliography{pd_ag_abun}

\begin{thebibliography}{96}
\expandafter\ifx\csname natexlab\endcsname\relax\def\natexlab#1{#1}\fi

\bibitem[{{Alonso} {et~al.}(1995){Alonso}, {Arribas}, \&
  {Martinez-Roger}}]{Alonso1995}
{Alonso}, A., {Arribas}, S., \& {Martinez-Roger}, C. 1995, \aap, 297, 197

\bibitem[{{Alonso} {et~al.}(1996){Alonso}, {Arribas}, \&
  {Martinez-Roger}}]{Alonso1996}
{Alonso}, A., {Arribas}, S., \& {Martinez-Roger}, C. 1996, \aap, 313, 873

\bibitem[{{Anstee} \& {O'Mara}(1991)}]{Anstee1991}
{Anstee}, S.~D. \& {O'Mara}, B.~J. 1991, \mnras, 253, 549

\bibitem[{{Anstee} \& {O'Mara}(1995)}]{Anstee1995}
{Anstee}, S.~D. \& {O'Mara}, B.~J. 1995, \mnras, 276, 859

\bibitem[{{Arce} \& {Goodman}(1999)}]{Arce1999}
{Arce}, H.~G. \& {Goodman}, A.~A. 1999, \apjl, 512, L135

\bibitem[{{Arcones} \& {Montes}(2011)}]{Arcones2011}
{Arcones}, A. \& {Montes}, F. 2011, \apj, 731, 5

\bibitem[{{Argast} {et~al.}(2004){Argast}, {Samland}, {Thielemann}, \&
  {Qian}}]{Argast2004}
{Argast}, D., {Samland}, M., {Thielemann}, F.-K., \& {Qian}, Y.-Z. 2004, \aap,
  416, 997

\bibitem[{{Arlandini} {et~al.}(1999){Arlandini}, {K{\"a}ppeler}, {Wisshak},
  {Gallino}, {Lugaro}, {Busso}, \& {Straniero}}]{Arlandini1999}
{Arlandini}, C., {K{\"a}ppeler}, F., {Wisshak}, K., {et~al.} 1999, \apj, 525,
  886

\bibitem[{{Arnould} \& {Goriely}(2003)}]{Arnould2003}
{Arnould}, M. \& {Goriely}, S. 2003, \physrep, 384, 1

\bibitem[{{Bagnulo} {et~al.}(2003){Bagnulo}, {Jehin}, {Ledoux}, {Cabanac},
  {Melo}, {Gilmozzi}, \& {The ESO Paranal Science Operations
  Team}}]{Bagnulo2003}
{Bagnulo}, S., {Jehin}, E., {Ledoux}, C., {et~al.} 2003, The Messenger, 114, 10

\bibitem[{{Bensby} {et~al.}(2003){Bensby}, {Feltzing}, \&
  {Lundstr{\"o}m}}]{Bensby2003}
{Bensby}, T., {Feltzing}, S., \& {Lundstr{\"o}m}, I. 2003, \aap, 410, 527

\bibitem[{{Bensby} {et~al.}(2014){Bensby}, {Feltzing}, \& {Oey}}]{Bensby2014}
{Bensby}, T., {Feltzing}, S., \& {Oey}, M.~S. 2014, \aap, 562, A71

\bibitem[{{Bisterzo} {et~al.}(2010){Bisterzo}, {Gallino}, {Straniero},
  {Cristallo}, \& {K{\"a}ppeler}}]{Bisterzo2010}
{Bisterzo}, S., {Gallino}, R., {Straniero}, O., {Cristallo}, S., \&
  {K{\"a}ppeler}, F. 2010, \mnras, 404, 1529

\bibitem[{{Bisterzo} {et~al.}(2011){Bisterzo}, {Gallino}, {Straniero},
  {Cristallo}, \& {K{\"a}ppeler}}]{Bisterzo2011}
{Bisterzo}, S., {Gallino}, R., {Straniero}, O., {Cristallo}, S., \&
  {K{\"a}ppeler}, F. 2011, \mnras, 418, 284

\bibitem[{{Boesgaard} \& {Novicki}(2006)}]{Boesgaard2006}
{Boesgaard}, A.~M. \& {Novicki}, M.~C. 2006, \apj, 641, 1122

\bibitem[{{Boesgaard} {et~al.}(2011){Boesgaard}, {Rich}, {Levesque}, \&
  {Bowler}}]{Boesgaard2011}
{Boesgaard}, A.~M., {Rich}, J.~A., {Levesque}, E.~M., \& {Bowler}, B.~P. 2011,
  \apj, 743, 140

\bibitem[{{Bonifacio} {et~al.}(2000){Bonifacio}, {Monai}, \&
  {Beers}}]{Bonifacio2000}
{Bonifacio}, P., {Monai}, S., \& {Beers}, T.~C. 2000, \aj, 120, 2065

\bibitem[{{Burbidge} {et~al.}(1957){Burbidge}, {Burbidge}, {Fowler}, \&
  {Hoyle}}]{Burbidge1957}
{Burbidge}, E.~M., {Burbidge}, G.~R., {Fowler}, W.~A., \& {Hoyle}, F. 1957,
  Reviews of Modern Physics, 29, 547

\bibitem[{{Burris} {et~al.}(2000){Burris}, {Pilachowski}, {Armandroff},
  {Sneden}, {Cowan}, \& {Roe}}]{Burris2000}
{Burris}, D.~L., {Pilachowski}, C.~A., {Armandroff}, T.~E., {et~al.} 2000,
  \apj, 544, 302

\bibitem[{{Busso} {et~al.}(1999){Busso}, {Gallino}, \&
  {Wasserburg}}]{Busso1999}
{Busso}, M., {Gallino}, R., \& {Wasserburg}, G.~J. 1999, \araa, 37, 239

\bibitem[{{Cameron}(1957)}]{Cameron1957}
{Cameron}, A.~G.~W. 1957, \pasp, 69, 201

\bibitem[{{Cameron}(2001)}]{Cameron2001}
{Cameron}, A.~G.~W. 2001, \apj, 562, 456

\bibitem[{{Canuto} \& {Mazzitelli}(1991)}]{Canuto1991}
{Canuto}, V.~M. \& {Mazzitelli}, I. 1991, \apj, 370, 295

\bibitem[{{Canuto} \& {Mazzitelli}(1992)}]{Canuto1992}
{Canuto}, V.~M. \& {Mazzitelli}, I. 1992, \apj, 389, 724

\bibitem[{{Chen} {et~al.}(2001){Chen}, {Nissen}, {Benoni}, \&
  {Zhao}}]{Chen2001}
{Chen}, Y.~Q., {Nissen}, P.~E., {Benoni}, T., \& {Zhao}, G. 2001, \aap, 371,
  943

\bibitem[{{Chen} {et~al.}(2000){Chen}, {Nissen}, {Zhao}, {Zhang}, \&
  {Benoni}}]{Chen2000}
{Chen}, Y.~Q., {Nissen}, P.~E., {Zhao}, G., {Zhang}, H.~W., \& {Benoni}, T.
  2000, \aaps, 141, 491

\bibitem[{{Clayton} \& {Rassbach}(1967)}]{Clayton1967}
{Clayton}, D.~D. \& {Rassbach}, M.~E. 1967, \apj, 148, 69

\bibitem[{{Crawford} {et~al.}(1998){Crawford}, {Sneden}, {King}, {Boesgaard},
  \& {Deliyannis}}]{Crawford1998}
{Crawford}, J.~L., {Sneden}, C., {King}, J.~R., {Boesgaard}, A.~M., \&
  {Deliyannis}, C.~P. 1998, \aj, 116, 2489

\bibitem[{{Cristallo} {et~al.}(2009){Cristallo}, {Straniero}, {Gallino},
  {Piersanti}, {Dom{\'{\i}}nguez}, \& {Lederer}}]{Cristallo2009}
{Cristallo}, S., {Straniero}, O., {Gallino}, R., {et~al.} 2009, \apj, 696, 797

\bibitem[{{Dehnen} \& {Binney}(1998)}]{Dehnen1998}
{Dehnen}, W. \& {Binney}, J.~J. 1998, \mnras, 298, 387

\bibitem[{{Dekker} {et~al.}(2000){Dekker}, {D'Odorico}, {Kaufer}, {Delabre}, \&
  {Kotzlowski}}]{Dekker2000}
{Dekker}, H., {D'Odorico}, S., {Kaufer}, A., {Delabre}, B., \& {Kotzlowski}, H.
  2000, in Society of Photo-Optical Instrumentation Engineers (SPIE) Conference
  Series, Vol. 4008, Society of Photo-Optical Instrumentation Engineers (SPIE)
  Conference Series, ed. {M.~Iye \& A.~F.~Moorwood}, 534--545

\bibitem[{{Farouqi} {et~al.}(2009){Farouqi}, {Kratz}, {Mashonkina}, {Pfeiffer},
  {Cowan}, {Thielemann}, \& {Truran}}]{Farouqi2009}
{Farouqi}, K., {Kratz}, K.-L., {Mashonkina}, L.~I., {et~al.} 2009, \apjl, 694,
  L49

\bibitem[{{Farouqi} {et~al.}(2010){Farouqi}, {Kratz}, {Pfeiffer}, {Rauscher},
  {Thielemann}, \& {Truran}}]{Farouqi2010}
{Farouqi}, K., {Kratz}, K.-L., {Pfeiffer}, B., {et~al.} 2010, \apj, 712, 1359

\bibitem[{{Fran{\c c}ois} {et~al.}(2007){Fran{\c c}ois}, {Depagne}, {Hill},
  {Spite}, {Spite}, {Plez}, {Beers}, {Andersen}, {James}, {Barbuy}, {Cayrel},
  {Bonifacio}, {Molaro}, {Nordstr{\"o}m}, \& {Primas}}]{Francois2007}
{Fran{\c c}ois}, P., {Depagne}, E., {Hill}, V., {et~al.} 2007, \aap, 476, 935

\bibitem[{{Freiburghaus} {et~al.}(1999){Freiburghaus}, {Rosswog}, \&
  {Thielemann}}]{Freiburghaus1999}
{Freiburghaus}, C., {Rosswog}, S., \& {Thielemann}, F.-K. 1999, \apjl, 525,
  L121

\bibitem[{{Frischknecht} {et~al.}(2012){Frischknecht}, {Hirschi}, \&
  {Thielemann}}]{Frischknecht2012}
{Frischknecht}, U., {Hirschi}, R., \& {Thielemann}, F.-K. 2012, \aap, 538, L2

\bibitem[{{Fuhrmann}(1998)}]{Fuhrmann1998}
{Fuhrmann}, K. 1998, \aap, 338, 161

\bibitem[{{Gallino} {et~al.}(1998){Gallino}, {Arlandini}, {Busso}, {Lugaro},
  {Travaglio}, {Straniero}, {Chieffi}, \& {Limongi}}]{Gallino1998}
{Gallino}, R., {Arlandini}, C., {Busso}, M., {et~al.} 1998, \apj, 497, 388

\bibitem[{{Gehren} {et~al.}(2006){Gehren}, {Shi}, {Zhang}, {Zhao}, \&
  {Korn}}]{Gehren2006}
{Gehren}, T., {Shi}, J.~R., {Zhang}, H.~W., {Zhao}, G., \& {Korn}, A.~J. 2006,
  \aap, 451, 1065

\bibitem[{{Goldberg} {et~al.}(2002){Goldberg}, {Mazeh}, {Latham}, {Stefanik},
  {Carney}, \& {Laird}}]{Goldberg2002}
{Goldberg}, D., {Mazeh}, T., {Latham}, D.~W., {et~al.} 2002, \aj, 124, 1132

\bibitem[{{Goriely} {et~al.}(2011){Goriely}, {Bauswein}, \&
  {Janka}}]{Goriely2011}
{Goriely}, S., {Bauswein}, A., \& {Janka}, H.-T. 2011, \apjl, 738, L32

\bibitem[{{Goriely} {et~al.}(2005){Goriely}, {Demetriou}, {Janka}, {Pearson},
  \& {Samyn}}]{Goriely2005}
{Goriely}, S., {Demetriou}, P., {Janka}, H.-T., {Pearson}, J.~M., \& {Samyn},
  M. 2005, Nuclear Physics A, 758, 587

\bibitem[{{Gratton} {et~al.}(2003){Gratton}, {Carretta}, {Claudi}, {Lucatello},
  \& {Barbieri}}]{Gratton2003}
{Gratton}, R.~G., {Carretta}, E., {Claudi}, R., {Lucatello}, S., \& {Barbieri},
  M. 2003, \aap, 404, 187

\bibitem[{{Grupp}(2004)}]{Grupp2004}
{Grupp}, F. 2004, \aap, 420, 289

\bibitem[{{Grupp} {et~al.}(2009){Grupp}, {Kurucz}, \& {Tan}}]{Grupp2009}
{Grupp}, F., {Kurucz}, R.~L., \& {Tan}, K. 2009, \aap, 503, 177

\bibitem[{{Hansen} \& {Primas}(2011)}]{Hansen2011}
{Hansen}, C.~J. \& {Primas}, F. 2011, \aap, 525, L5

\bibitem[{{Hansen} {et~al.}(2012){Hansen}, {Primas}, {Hartman}, {Kratz},
  {Wanajo}, {Leibundgut}, {Farouqi}, {Hallmann}, {Christlieb}, \&
  {Nilsson}}]{Hansen2012}
{Hansen}, C.~J., {Primas}, F., {Hartman}, H., {et~al.} 2012, \aap, 545, A31

\bibitem[{{Hauck} \& {Mermilliod}(1998)}]{Hauck1998}
{Hauck}, B. \& {Mermilliod}, M. 1998, \aaps, 129, 431

\bibitem[{{Hill} {et~al.}(2002){Hill}, {Plez}, {Cayrel}, {Beers},
  {Nordstr{\"o}m}, {Andersen}, {Spite}, {Spite}, {Barbuy}, {Bonifacio},
  {Depagne}, {Fran{\c c}ois}, \& {Primas}}]{Hill2002}
{Hill}, V., {Plez}, B., {Cayrel}, R., {et~al.} 2002, \aap, 387, 560

\bibitem[{{H{\o}g} {et~al.}(2000){H{\o}g}, {Fabricius}, {Makarov}, {Urban},
  {Corbin}, {Wycoff}, {Bastian}, {Schwekendiek}, \& {Wicenec}}]{Hog2000}
{H{\o}g}, E., {Fabricius}, C., {Makarov}, V.~V., {et~al.} 2000, \aap, 355, L27

\bibitem[{{Honda} {et~al.}(2006){Honda}, {Aoki}, {Ishimaru}, {Wanajo}, \&
  {Ryan}}]{Honda2006}
{Honda}, S., {Aoki}, W., {Ishimaru}, Y., {Wanajo}, S., \& {Ryan}, S.~G. 2006,
  \apj, 643, 1180

\bibitem[{{Honda} {et~al.}(2004){Honda}, {Aoki}, {Kajino}, {Ando}, {Beers},
  {Izumiura}, {Sadakane}, \& {Takada-Hidai}}]{Honda2004}
{Honda}, S., {Aoki}, W., {Kajino}, T., {et~al.} 2004, \apj, 607, 474

\bibitem[{{Ishigaki} {et~al.}(2012){Ishigaki}, {Chiba}, \&
  {Aoki}}]{Ishigaki2012}
{Ishigaki}, M.~N., {Chiba}, M., \& {Aoki}, W. 2012, \apj, 753, 64

\bibitem[{{Johnson} \& {Soderblom}(1987)}]{Johnson1987}
{Johnson}, D.~R.~H. \& {Soderblom}, D.~R. 1987, \aj, 93, 864

\bibitem[{{Johnson} \& {Bolte}(2002)}]{Johnson2002}
{Johnson}, J.~A. \& {Bolte}, M. 2002, \apj, 579, 616

\bibitem[{{Just} {et~al.}(2015){Just}, {Bauswein}, {Pulpillo}, {Goriely}, \&
  {Janka}}]{Just2015}
{Just}, O., {Bauswein}, A., {Pulpillo}, R.~A., {Goriely}, S., \& {Janka}, H.-T.
  2015, \mnras, 448, 541

\bibitem[{{K{\"a}ppeler} {et~al.}(1989){K{\"a}ppeler}, {Beer}, \&
  {Wisshak}}]{Kappeler1989}
{K{\"a}ppeler}, F., {Beer}, H., \& {Wisshak}, K. 1989, Reports on Progress in
  Physics, 52, 945

\bibitem[{{K{\"a}ppeler} {et~al.}(2011){K{\"a}ppeler}, {Gallino}, {Bisterzo},
  \& {Aoki}}]{Kappeler2011}
{K{\"a}ppeler}, F., {Gallino}, R., {Bisterzo}, S., \& {Aoki}, W. 2011, Reviews
  of Modern Physics, 83, 157

\bibitem[{{Karakas} {et~al.}(2009){Karakas}, {van Raai}, {Lugaro}, {Sterling},
  \& {Dinerstein}}]{Karakas2009}
{Karakas}, A.~I., {van Raai}, M.~A., {Lugaro}, M., {Sterling}, N.~C., \&
  {Dinerstein}, H.~L. 2009, \apj, 690, 1130

\bibitem[{{Korn} {et~al.}(2003){Korn}, {Shi}, \& {Gehren}}]{Korn2003}
{Korn}, A.~J., {Shi}, J., \& {Gehren}, T. 2003, \aap, 407, 691

\bibitem[{{Korobkin} {et~al.}(2012){Korobkin}, {Rosswog}, {Arcones}, \&
  {Winteler}}]{Korobkin2012}
{Korobkin}, O., {Rosswog}, S., {Arcones}, A., \& {Winteler}, C. 2012, \mnras,
  426, 1940

\bibitem[{{Kratz} {et~al.}(2007){Kratz}, {Farouqi}, {Pfeiffer}, {Truran},
  {Sneden}, \& {Cowan}}]{Kratz2007}
{Kratz}, K.-L., {Farouqi}, K., {Pfeiffer}, B., {et~al.} 2007, \apj, 662, 39

\bibitem[{{Kurucz}(2005)}]{Kurucz2005}
{Kurucz}, R.~L. 2005, Memorie della Societa Astronomica Italiana Supplementi,
  8, 189

\bibitem[{{Lind} {et~al.}(2012){Lind}, {Bergemann}, \& {Asplund}}]{Lind2012}
{Lind}, K., {Bergemann}, M., \& {Asplund}, M. 2012, \mnras, 427, 50

\bibitem[{{Lodders} {et~al.}(2009){Lodders}, {Palme}, \& {Gail}}]{Lodders2009}
{Lodders}, K., {Palme}, H., \& {Gail}, H.-P. 2009, Landolt B{\"o}rnstein, 44

\bibitem[{{Mashonkina} {et~al.}(2011){Mashonkina}, {Gehren}, {Shi}, {Korn}, \&
  {Grupp}}]{Mashonkina2011}
{Mashonkina}, L., {Gehren}, T., {Shi}, J.-R., {Korn}, A.~J., \& {Grupp}, F.
  2011, \aap, 528, A87

\bibitem[{{McLaughlin} \& {Surman}(2005)}]{McLaughlin2005}
{McLaughlin}, G.~C. \& {Surman}, R. 2005, Nuclear Physics A, 758, 189

\bibitem[{{Montes} {et~al.}(2007){Montes}, {Beers}, {Cowan}, {Elliot},
  {Farouqi}, {Gallino}, {Heil}, {Kratz}, {Pfeiffer}, {Pignatari}, \&
  {Schatz}}]{Montes2007}
{Montes}, F., {Beers}, T.~C., {Cowan}, J., {et~al.} 2007, \apj, 671, 1685

\bibitem[{{Navarro} {et~al.}(2011){Navarro}, {Abadi}, {Venn}, {Freeman}, \&
  {Anguiano}}]{Navarro2011}
{Navarro}, J.~F., {Abadi}, M.~G., {Venn}, K.~A., {Freeman}, K.~C., \&
  {Anguiano}, B. 2011, \mnras, 412, 1203

\bibitem[{{Nissen}(2013)}]{Nissen2013}
{Nissen}, P.~E. 2013, \aap, 552, A73

\bibitem[{{Noguchi} {et~al.}(2002){Noguchi}, {Aoki}, {Kawanomoto}, {Ando},
  {Honda}, {Izumiura}, {Kambe}, {Okita}, {Sadakane}, {Sato}, {Tajitsu},
  {Takada-Hidai}, {Tanaka}, {Watanabe}, \& {Yoshida}}]{Noguchi2002}
{Noguchi}, K., {Aoki}, W., {Kawanomoto}, S., {et~al.} 2002, \pasj, 54, 855

\bibitem[{{Perego} {et~al.}(2014){Perego}, {Rosswog}, {Cabez{\'o}n},
  {Korobkin}, {K{\"a}ppeli}, {Arcones}, \& {Liebend{\"o}rfer}}]{Perego2014}
{Perego}, A., {Rosswog}, S., {Cabez{\'o}n}, R.~M., {et~al.} 2014, \mnras, 443,
  3134

\bibitem[{{Peterson}(2013)}]{Peterson2013}
{Peterson}, R.~C. 2013, \apjl, 768, L13

\bibitem[{{Pignatari} {et~al.}(2010){Pignatari}, {Gallino}, {Heil}, {Wiescher},
  {K{\"a}ppeler}, {Herwig}, \& {Bisterzo}}]{Pignatari2010}
{Pignatari}, M., {Gallino}, R., {Heil}, M., {et~al.} 2010, \apj, 710, 1557

\bibitem[{{Pignatari} {et~al.}(2013){Pignatari}, {Hirschi}, {Wiescher},
  {Gallino}, {Bennett}, {Beard}, {Fryer}, {Herwig}, {Rockefeller}, \&
  {Timmes}}]{Pignatari2013}
{Pignatari}, M., {Hirschi}, R., {Wiescher}, M., {et~al.} 2013, \apj, 762, 31

\bibitem[{{Ram{\'{\i}}rez} \& {Mel{\'e}ndez}(2004)}]{Ramirez2004}
{Ram{\'{\i}}rez}, I. \& {Mel{\'e}ndez}, J. 2004, \apj, 609, 417

\bibitem[{{Reddy} {et~al.}(2003){Reddy}, {Tomkin}, {Lambert}, \& {Allende
  Prieto}}]{Reddy2003}
{Reddy}, B.~E., {Tomkin}, J., {Lambert}, D.~L., \& {Allende Prieto}, C. 2003,
  \mnras, 340, 304

\bibitem[{{Reetz}(1991)}]{Reetz1991}
{Reetz}, J.~K. 1991, {Diploma Thesis, Universit\"{a}t M\"{u}nchen}

\bibitem[{{Ross} \& {Aller}(1972)}]{Ross1972}
{Ross}, J.~E. \& {Aller}, L.~H. 1972, \solphys, 25, 30

\bibitem[{{Rosswog} {et~al.}(1999){Rosswog}, {Liebend{\"o}rfer}, {Thielemann},
  {Davies}, {Benz}, \& {Piran}}]{Rosswog1999}
{Rosswog}, S., {Liebend{\"o}rfer}, M., {Thielemann}, F.-K., {et~al.} 1999,
  \aap, 341, 499

\bibitem[{{Schlegel} {et~al.}(1998){Schlegel}, {Finkbeiner}, \&
  {Davis}}]{Schlegel1998}
{Schlegel}, D.~J., {Finkbeiner}, D.~P., \& {Davis}, M. 1998, \apj, 500, 525

\bibitem[{{Simmerer} {et~al.}(2004){Simmerer}, {Sneden}, {Cowan}, {Collier},
  {Woolf}, \& {Lawler}}]{Simmerer2004}
{Simmerer}, J., {Sneden}, C., {Cowan}, J.~J., {et~al.} 2004, \apj, 617, 1091

\bibitem[{{Sneden} {et~al.}(2008){Sneden}, {Cowan}, \& {Gallino}}]{Sneden2008}
{Sneden}, C., {Cowan}, J.~J., \& {Gallino}, R. 2008, \araa, 46, 241

\bibitem[{{Sneden} {et~al.}(2003){Sneden}, {Cowan}, {Lawler}, {Ivans},
  {Burles}, {Beers}, {Primas}, {Hill}, {Truran}, {Fuller}, {Pfeiffer}, \&
  {Kratz}}]{Sneden2003}
{Sneden}, C., {Cowan}, J.~J., {Lawler}, J.~E., {et~al.} 2003, \apj, 591, 936

\bibitem[{{Straniero} {et~al.}(1997){Straniero}, {Chieffi}, {Limongi}, {Busso},
  {Gallino}, \& {Arlandini}}]{Straniero1997}
{Straniero}, O., {Chieffi}, A., {Limongi}, M., {et~al.} 1997, \apj, 478, 332

\bibitem[{{Tan} {et~al.}(2009){Tan}, {Shi}, \& {Zhao}}]{Tan2009}
{Tan}, K.~F., {Shi}, J.~R., \& {Zhao}, G. 2009, \mnras, 392, 205

\bibitem[{{Travaglio} {et~al.}(2004){Travaglio}, {Gallino}, {Arnone}, {Cowan},
  {Jordan}, \& {Sneden}}]{Travaglio2004}
{Travaglio}, C., {Gallino}, R., {Arnone}, E., {et~al.} 2004, \apj, 601, 864

\bibitem[{{Travaglio} {et~al.}(2001){Travaglio}, {Gallino}, {Busso}, \&
  {Gratton}}]{Travaglio2001}
{Travaglio}, C., {Gallino}, R., {Busso}, M., \& {Gratton}, R. 2001, \apj, 549,
  346

\bibitem[{{van Leeuwen}(2007)}]{vanLeeuwen2007}
{van Leeuwen}, F. 2007, \aap, 474, 653

\bibitem[{{Vogt} {et~al.}(1994){Vogt}, {Allen}, {Bigelow}, {Bresee}, {Brown},
  {Cantrall}, {Conrad}, {Couture}, {Delaney}, {Epps}, {Hilyard}, {Hilyard},
  {Horn}, {Jern}, {Kanto}, {Keane}, {Kibrick}, {Lewis}, {Osborne},
  {Pardeilhan}, {Pfister}, {Ricketts}, {Robinson}, {Stover}, {Tucker}, {Ward},
  \& {Wei}}]{Vogt1994}
{Vogt}, S.~S., {Allen}, S.~L., {Bigelow}, B.~C., {et~al.} 1994, in Society of
  Photo-Optical Instrumentation Engineers (SPIE) Conference Series, Vol. 2198,
  Instrumentation in Astronomy VIII, ed. D.~L. {Crawford} \& E.~R. {Craine},
  362

\bibitem[{{Wanajo} \& {Janka}(2012)}]{Wanajo2012}
{Wanajo}, S. \& {Janka}, H.-T. 2012, \apj, 746, 180

\bibitem[{{Wanajo} {et~al.}(2003){Wanajo}, {Tamamura}, {Itoh}, {Nomoto},
  {Ishimaru}, {Beers}, \& {Nozawa}}]{Wanajo2003}
{Wanajo}, S., {Tamamura}, M., {Itoh}, N., {et~al.} 2003, \apj, 593, 968

\bibitem[{{Wasserburg} \& {Qian}(2000)}]{Wasserburg2000}
{Wasserburg}, G.~J. \& {Qian}, Y.-Z. 2000, \apjl, 529, L21

\bibitem[{{Woosley} {et~al.}(1997){Woosley}, {Hoffman}, {Timmes}, {Weaver}, \&
  {Thielemann}}]{Woosley1997}
{Woosley}, S.~E., {Hoffman}, R.~D., {Timmes}, F.~X., {Weaver}, T.~A., \&
  {Thielemann}, F.-K. 1997, Nuclear Physics A, 621, 445

\bibitem[{{Yi} {et~al.}(2003){Yi}, {Kim}, \& {Demarque}}]{Yi2003}
{Yi}, S.~K., {Kim}, Y.-C., \& {Demarque}, P. 2003, \apjs, 144, 259

\bibitem[{{Zhao} \& {Magain}(1990)}]{Zhao1990}
{Zhao}, G. \& {Magain}, P. 1990, \aap, 238, 242

\end{thebibliography}

\end{document}